\documentclass[useAMS,usenatbib]{mn2e}

\input{psfig.sty}
\usepackage{graphicx}
\usepackage{epsfig}
\usepackage{amssymb}
\usepackage{amsmath}
\usepackage{amsfonts}
\usepackage{txfonts}
\usepackage{multirow}

\title[CUBEP$^3$M]{High Performance P$^{3}$M N-body code: CUBEP$^3$M}
\author[Joachim Harnois-D\'{e}raps et al.]{Joachim Harnois-D\'{e}raps$^{1,2}$ 
\thanks{E-mail: jharno@cita.utoronto.ca},  Ue-Li Pen$^{1}$, 
Ilian T. Iliev$^{3}$, Hugh Merz$^{4}$, \newauthor
J.D. Emberson$^{1,5}$ and Vincent Desjacques$^{6}$\\
$^{1}$Canadian Institute for Theoretical Astrophysics, University of
Toronto, M5S 3H8, Ontario, Canada\\
$^{2}$Department of Physics, University of Toronto, M5S 1A7, Ontario,  Canada\\
$^{3}$Astronomy Centre, Department of Physics and Astronomy, Pevensey II Building, University of Sussex, BN1 9QH, Brighton, United Kingdom\\
$^{4}$SHARCNET, Laurentian University, P3E 2C6, Ontario, Canada\\
$^{5}$Department of Astronomy and Astrophysics, University of Toronto, M5S 3H4, Ontario, Canada\\
$^{6}$Universit\'{e} de Gen\`{e}ve and Center for Astroparticle Physics, 24 Quai Ernest Ansermet, 1211 Gen\`{e}ve 4, Switzerland}

\begin{document}

%\date{Accepted 1988 December 15. Received 1988 December 14; in original form 1988 October 11}
\date{\today}

\pagerange{\pageref{firstpage}--\pageref{lastpage}} \pubyear{2011}

\maketitle

\label{firstpage}

\begin{abstract}
This paper presents {\small CUBEP$^3$M}, a publicly-available high performance cosmological N-body code
and describes many utilities and extensions that have been added to the standard package. 
These include a memory-light runtime SO halo finder,
a non-Gaussian initial conditions generator, and a system of unique particle identification.
{\small CUBEP$^3$M} is fast, its accuracy is tuneable to optimize speed or memory, 
and has been run on more than $27,000$ cores, achieving within a factor of two of ideal weak scaling
even at this problem size.
The code can be run in an extra-lean mode where the peak memory imprint for large runs is as low as $37$ bytes per particles, 
which is almost two times leaner than other widely used N-body codes. 
However, load imbalances can increase this requirement by a factor of two,
such that fast configurations with all the utilities enabled and load imbalances factored in require between $70$ and $120$ bytes per particles.
{\small CUBEP$^3$M} is well designed to study large scales cosmological systems, where imbalances are not too large and adaptive time-stepping not essential.
It has already been used for a broad number of science applications that 
require either large samples of non-linear realizations or very large dark matter N-body simulations,
including  cosmological reionization, halo formation, baryonic acoustic oscillations, weak lensing or non-Gaussian statistics.
We discuss the structure, the accuracy, known systematic effects and the scaling performance
of the code and its utilities, when applicable.
\end{abstract}


%%%%%%%%%%%%%%%

\section{Introduction}

Many physical and astrophysical systems are subject to non-linear dynamics
and rely on N-body simulations to describe the evolution of bodies. 
One of the main field of application is the modelling of large scale structures, 
which are driven by the sole force of gravity. Recent observations of the 
cosmic microwave background \citep{2009ApJS..180..330K,2011ApJS..192...18K}, of galaxy clustering 
\citep{2000AJ....120.1579Y, 2003astro.ph..6581C, 2009arXiv0902.4680S, 
2010MNRAS.401.1429D} of weak gravitational lensing \citep{2012AAS...21913001H, 2009ApJ...703.2232S}
and of supernovae redshift-distance relations all point towards a standard 
model of cosmology, in which dark energy and collision-less dark matter occupy 
more than 95 per cent of the total energy density of the universe. In such a 
paradigm, pure N-body codes are perfectly suited to describe the dynamics, as 
long as baryonic physics are not very important, or at least we understand how 
the baryonic fluid feeds back on the dark matter structure. The next generation 
of measurements aims at constraining the cosmological parameters at the per cent 
level, and the theoretical understanding of the non-linear dynamics that govern 
structure formation heavily relies on numerical simulations. 

For instance, a measurement of the baryonic acoustic oscillation (BAO) dilation 
scale can provide tight constraints on the dark energy equation of state 
\citep{Eisenstein:2005su,2006PhRvD..74l3507T, 2007MNRAS.381.1053P,2009arXiv0902.4680S}. 
The most optimal estimates of the uncertainty requires the 
knowledge of the matter power spectrum covariance matrix, which is only accurate 
when measured from a large sample of N-body simulations \citep{2005MNRAS.360L..82R, 
2009ApJ...700..479T, 2011ApJ...726....7T}. For the same reasons, the most accurate 
estimates of weak gravitational lensing signals are obtained by propagating photons 
in past light cones that are extracted from simulated density fields 
\citep{2003ApJ...592..699V, 2009ApJ...701..945S, 2009A&A...499...31H}.
Another area where large-scale N-body simulations have been 
instrumental  in recent years are in simulations of early cosmic structures and reionization 
\citep[e.g.][]{2006MNRAS.369.1625I,2007ApJ...654...12Z,2007ApJ...671....1T,
2012MNRAS.423.2222I}. The reionization process is primarily driven by low-mass 
galaxies, which for both observational and theoretical reasons, need to be resolved 
in fairly large volumes, which demands simulations with a large dynamical range.   
%{\bf (Ilian, could you say something about reionization here? Just general references as to why one needs N-body codes to perform 
%accurate predictions/forecasts, etc.)}

The basic problem that is addressed with N-body codes is the time evolution of an ensemble of $N$ particles
that is subject to gravitational attraction. 
The brute force calculation requires $O(N^{2})$ operations, a cost that 
exceeds the memory and speed of current machines for large problems.
Solving the problem  on a mesh \citep{1981csup.book.....H} reduces to $O(N\mbox{log}N)$ the number of operations,
as it is possible to solve for the particle-mesh (PM) interaction with fast Fourier transforms techniques that rely on high performance libraries such as {\small FFTW} \citep{FFTW3}.

With the advent of large computing facilities, parallel computations have now become 
common practice, and N-body codes have evolved both in performance and complexity. 
Many have opted for  `tree' algorithms, including {\small GADGET} \citep{2001NewA....6...79S, 2005MNRAS.364.1105S,2012MNRAS.426.2046A}, 
{\small PMT} \citep{1995ApJS...98..355X}, {\small GOTPM} \citep{2004NewA....9..111D}, or adaptive P$^3$M codes like Hydra \citep{1995ApJ...452..797C} or {\small RAMSES} \citep{2010ascl.soft11007T}, 
in which the local resolution increases with the density of the matter field. 
These often have the advantage to balance the work load across the computing units, which enable relatively fast calculations even in high density regions. 
The drawback is a significant loss in speed, and a general increase in memory consumption, which can be only partly recovered by turning off the tree algorithm. 
The same reasoning applies to mesh-refined codes  \citep{1991ApJ...368L..23C}, 
which in the end are not designed to perform fast PM calculations on large scales. 

Although such codes are essential to study systems that are spatially heterogeneous like individual clusters or haloes, AGNs or other compact objects,
many applications are interested in studying large cosmological volumes in which the matter distribution is rather homogeneous.
In addition, statistical analyses typically require running a large number of simulations;
in such systems, the load balancing algorithm is not necessary, and the effort
can thereby be put towards speed, memory compactness and scalability.  
{\small PMFAST} (\citet{2005NewA...10..393M}, MPT hereafter) was one of the first code designed specifically to optimize the PM algorithm,
both in terms of speed and memory usage, and uses a two-level mesh algorithm based on the gravity solver of \citet{2003AAS...203.9703T}.
The long range gravitational force is computed on a  grid $4^{3}$ times coarser, such as to minimize the {\small MPI} communication time
and to fit in system's memory. The short range is computed locally on a finer mesh, and only the local sub-volume needs 
to be stored at a given time, allowing for {\small OPENMP} parallel computation.
This setup enables the code to evolve large cosmological systems both rapidly and accurately, even on relatively modest clusters.
One of the main advantages of {\small PMFAST} over other PM codes is that the number of large arrays is minimized,
and the global {\small MPI} communications are cut down to the minimum: for passing particles at the beginning of each time step,
and  for computing the long range FFTs.

Since its first published version, {\small PMFAST} has evolved in many aspects. 
The first major improvement was to transform the volume decomposition in multi-node configurations 
from slabs to cubes. The problem with slabs is that they do not scale well to large runs: 
as the number of cells per dimension increases, the thickness of each slab shrinks rapidly,
until it reaches the hard limit of a single cell layer.  With this enhancement, the code name was changed to {\small CUBEPM}. Soon after, it incorporated particle-particle (pp) interactions at the sub-grid level, 
and was finally renamed {\small CUBEP$^3$M}. The public package now includes a significant number of new features: the pp force
can be extended to an arbitrary range, the size of the redshift jumps can be constrained for improved accuracy during the first time steps,
a runtime halo finder has been implemented, the expansion has also been generalized to include a redshift dependent equation of state for dark energy, there is a system of unique particle identification that can be switched on or off, and the initial condition generator has been generalized as to include non-Gaussian features.
 {\small PMFAST} was equipped with a multi-time stepping option that has not been tested on {\small CUBEP$^3$M} yet, but which is, in principle at least, still available. 
 
 The standard package also contains support for magneto-hydrodynamical gas evolution through a portable TVD-MHD module \citep{2003ApJS..149..447P} that scales up to thousands of cores as well (see \citet{2010arXiv1004.1680P} and footnote 4 in section \ref{sec:scaling}), and a coupling interface with the radiative transfer code C$^2$-ray \citep{2006NewA...11..374M}.
 {\small CUBEP$^3$M} is therefore one of the most competitive and versatile public N-body code, 
 and has been involved in a number of scientific applications over the last few years,
spanning the field of weak lensing \citep{Vafaei10, 2008MNRAS.388.1819L,  2009arXiv0905.0501D, 2010PhRvD..81l3015L, 2012MNRAS.421..832Y, 2012MNRAS.426.1262H},  BAO
 \citep{2011ApJ...728...35Z,  2012MNRAS.419.2949N, 2012MNRAS.423.2288H, 2012arXiv1205.4989H}, 
formation of early cosmic structures \citep{2008arXiv0806.2887I,2010arXiv1005.2502I},
observations of dark stars \citep{2010MNRAS.407L..74Z,2012MNRAS.422.2164I} and
reionization \citep{2012MNRAS.423.2222I,2012ApJ...750...20F,2011MNRAS.413.1353F,2012MNRAS.422..926M,2012MNRAS.424.1877D,2012MNRAS.424..762D}. 
Continuous efforts are being made to develop, extend and improve the code and each of its utilities, 
and we expect that this will pave the way to an increasing number of science projects. 
Notably, the fact that the force calculation is multi-layered makes the code extendible, 
and opens the possibility to run {\small CUBEP$^3$M} on hybrid CPU-GPU clusters.
It is thus important for the community to have access to a paper that describes the methodology, the accuracy and the performance of this public code. 

%\subsection{Simulation suites}
%\label{subsec:suites}

Since {\small CUBEP$^3$M} is not new, the existing  accuracy and systematic tests were performed by different groups, on different machines,
and with different geometric and parallelization configurations. It is not an ideal situation in which to quantify the performance, and each test must therefore be viewed as a
separate measurement that quantifies a specific aspect of the code. 
We have tried to keep to a minimum the number of such different test runs, and although the detailed numbers vary 
with the problem size and the machines, the general trends are rather universal.

Tests on improvements of the force calculations were performed 
on the Sunnyvale beowulf cluster at the Canadian Institute for Theoretical Astrophysics (CITA).
Each node contains 2 Quad Core Intel(R) Xeon(R) E5310 1.60GHz processors, 4GB of RAM,  a 40 GB disk and 2 gigE network interfaces. 
%These tests were performed on the same cluster, but with  different box sizes, starting redshifts and particle numbers.
Hereafter we refer to these simulation sets as the CITA128/256 configurations, 
which evolved $128^3$ and $256^3$ particles respectively in boxes of $100$, $150$ and $1000 h^{-1}\mbox{Mpc}$ (
we specify which parameters were used in each case).
Some complimentary test runs were also performed on the SciNet GPC cluster \citep{Scinet},
a system of IBM iDataPlex DX360M2 nodes equipped with Intel Xeon E5540 cores  running at 2.53GHz with 2GB of RAM per core.
For tests of the code accuracy, of the non-Gaussian initial conditions generator and of the run time halo finder algorithm, 
we used a third simulation configuration series that was run at the Texas Advanced Computing Centre (TACC) on Ranger, a SunBlade 
x6420 system with AMD x86\_64 Opteron Quad Core 2.3 GHz `Barcelona' processors and Infiniband networking.
These RANGER4000 simulations evolved $4000^{3}$ particles 
from $z=100$ to $z=0$ with a box side of $3.2 h^{-1}\mbox{Gpc}$
on 4000 cores. Finally, scaling tests were performed on the Curie and JUROPA systems, 
which are described in section \ref{sec:scaling}.

The paper is structured as follow: section \ref{sec:structure} reviews the structure and flow of the main code;
section \ref{sec:Poisson} describes how Poisson equation is solved on the two-mesh system;
we then present in section \ref{sec:scaling} the scaling of the code to very large problems;
section \ref{sec:accuracy}  discusses the accuracy and systematic effects of the code.
 We then describe in section \ref{sec:halo} the run-time halo finder, in section \ref{sec:extensions} various extensions
 to the default configuration, and conclude afterwards.

\section{Review of the Code Structure}
\label{sec:structure}

An optimal large scale N-body code must address many challenges: minimize the memory footprint to allow larger dynamical range,
minimize the passing of information across computing nodes, reduce and accelerate the memory accesses to the large scale arrays, 
make efficient use of high performance libraries to speed up standard calculations like Fourier transforms, etc.
In the realm of parallel programming, high efficiency  can be assessed when a high load is balanced across all processors
most of the time. In this section, we present the general strategies adopted to address these challenges\footnote{ 
Many originate directly from MPT and were preserved in {\small CUBEP$^3$M};
those will be briefly mentioned, and we shall refer the reader to the original {\small PMFAST} paper for greater details.}.
We start with a walkthrough the code flow, and briefly discuss some specific sections that depart from standard N-body codes,
while referring the reader to future sections for detailed discussions on selected topics.

As mentioned in the Introduction section, {\small CUBEP$^3$M} is a {\small FORTRAN90} 
N-body code that solves Poisson's equation on a two-level mesh, 
with sub-cell accuracy thanks to particle-particle interactions. 
The code has extensions that departs from this basic scheme, and
we shall come back to these later, but for the moment, we adopt the 
standard configuration. 

\begin{figure*}%[ht]
  \begin{center}
 \vspace{0cm}
 \hspace{-2cm}
% \hspace{-6cm}
 \centering
    \includegraphics[width=4.2in]{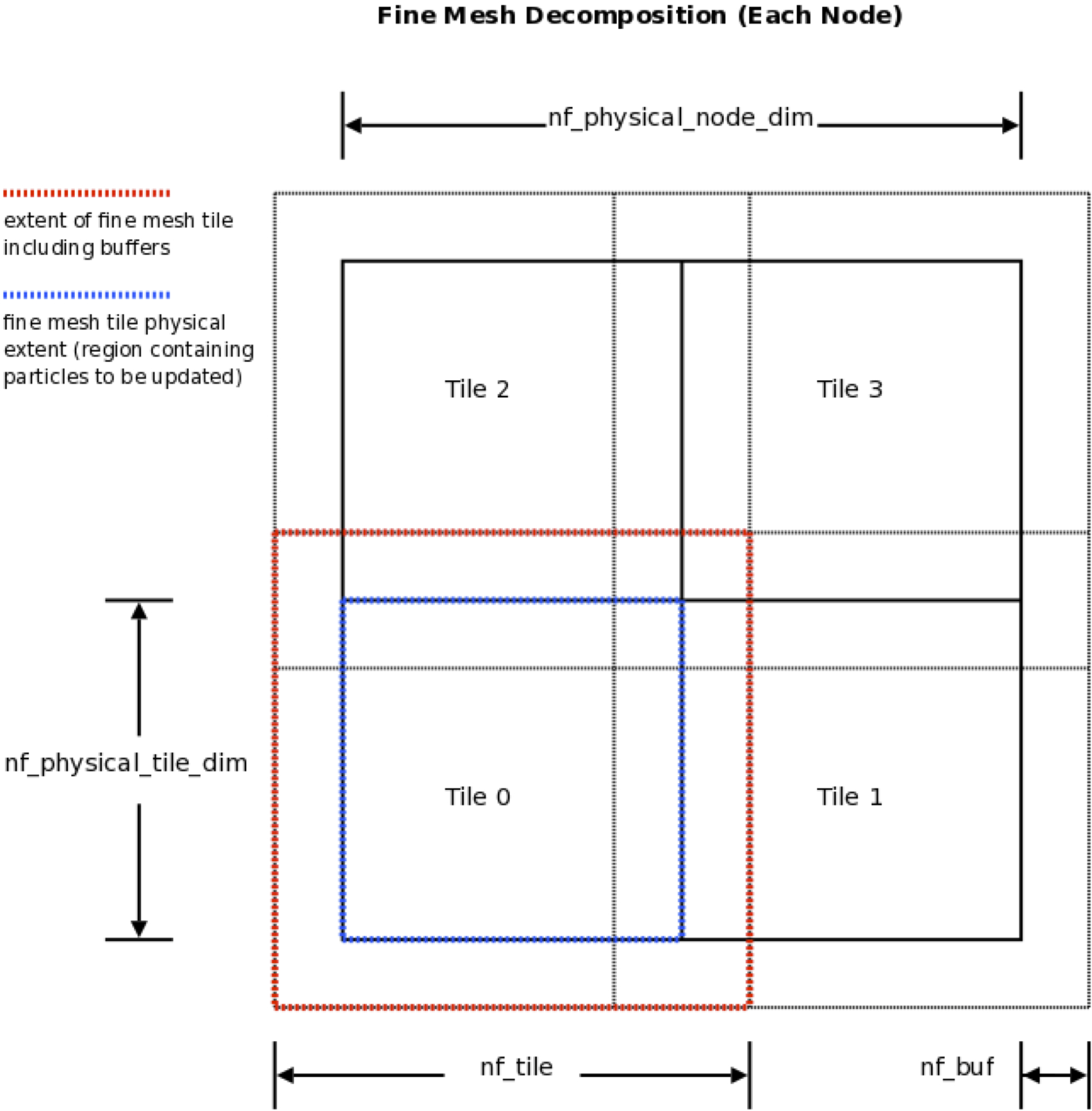}
\caption{Overall volume decomposition inside a node, simplified to a 2-dimensional side view. 
              In this example, there are two tiles per dimensions per node, for a total 8 tiles per node.
              Each tile works on a sub-volume that has {\tt nf\_physical\_tile\_dim} fine mesh cells per side,
              which is surrounded by a buffer region that is {\tt nf\_buf} (=24) fine cells thick. Operations on tiles are {\small OPENMP} threaded
              on each node.}
%\label{fig:volume_decomp_local}
%\end{center}
%\end{figure*}
%
%\begin{figure}%[ht]
  %\begin{center}
 %   \includegraphics[width=2.2in]{./Particle_list_decomp.ps}
%\caption{Overall volume decomposition across nodes. }
\label{fig:volume_decomp_global}
 \end{center}
\end{figure*}

The global simulation volume of first broken down in small cubical sub-volumes, each of which
is assigned to a distinct {\small MPI} task. Unless otherwise specified, we assume that each node works on one task to ease the reading,
even though a single node can hold two or more of such tasks if desired.
A  second level of cubical decomposition occurs inside each node, as 
depicted in Fig. \ref{fig:volume_decomp_global}:
the nodes sub-volumes are broken in a number of local volumes, the {\it tiles},
which are basically independent tasks that can be done in parallel.
Since the volumetric decomposition is cubical at both levels,
then both the number of tiles per nodes and the total number of nodes
must be perfect cubes (not necessarily the same).

The long range component of the gravity force is solved on the coarse grid, 
and is global in the sense that the calculations require knowledge about the full simulated volume,
involving {\small MPI} communications across nodes.
The short range force and the particle-particle interactions are computed on the fine grid,
one tile per processor at a time, in such a way as to allow for safe {\small OPENMP} threads.
To make this possible, the fine grid arrays are constructed such as to support parallel memory access. In practice, this is done by adding an additional dimension to the relevant arrays, such that each {\small CPU} accesses a unique memory location. The force matching between the two meshes is performed by introducing a cutoff length, $r_{c}$, in the definition of the two force kernels. The value of $r_{c}=16$ fine cells was found to balance the communication 
overhead between processes and the accuracy of the match between the two meshes. 
The default resolution ratio between the two grids are 1:4, although
the code is constructed as to support other ratios as well.
One would only need to modify the {\tt mesh\_scale} parameter, 
to feed force kernel files that are adjusted to this new ratio,
and to optimize the value of $r_{c}$.

The computation of the short range force requires each tile to store the fine grid density of a region that includes a buffer surface around the physical volume it is assigned. The thickness of this surface must be larger than $r_{c}$, and we find that a {\tt nf\_buff} $= 24$  cells deep buffer
is a good compromise between memory usage and accuracy (See Fig. \ref{fig:volume_decomp_global}).
This fully compensates for the coarse mesh calculations, whose CIC interpolation scheme 
reaches two coarse cells deep beyond the cutoff.

When it comes to finding haloes at run time, this buffer can create a problem, because a large object located close to the boundary can have a radius larger than the buffer zone, in which case it would be truncated and be assigned a wrong mass, centre of mass, etc. 
It could then be desirable to increase the buffer zone around each tile, at the cost of a loss of memory dedicated to the actual physical volume, and the code is designed to allow for such a flexibility.

Three important processes that speed up the code were already present in {\small PMFAST}: 
1) access to particles is accelerated with the use of linked lists, 
2) deletion of `ghost' particles in buffer zones is done at the same time as particles are passed to adjacent nodes, and
3) the global FFTs are performed with a slab decomposition of the volumes via a special file transfer interface, 
designed specifically to preserve a high processor load.

%\subsection{Memory foot-print and communication strategy}
%\label{subsec:memory}

Because the coarse grid arrays require $4^3$ times less memory per node, 
it does not contribute much to the total memory requirement, and the bulk of the foot-print is 
concentrated in a handful of fine grid arrays.
Some of these are only required for intermediate steps of the calculations, 
hence it is possible to hide therein many coarse grid arrays\footnote{ This memory recycling is done with `equivalence' statements in {\small FORTRAN90}.}.   
We present here the largest arrays used by the code:
\begin{enumerate}
\item{{\tt xv} stores the position and velocity of each particle} 
\item{{\tt ll} stores the linked-list that accelerate the access to particles in each coarse grid cell}
\item{{\tt send\_buf} and  {\tt recv\_buf} store the particles to be {\small MPI}-passed to the neighbouring sub-volumes}
\item{{\tt rho\_f} and {\tt cmplx\_rho\_f} store 
the local fine grid density  in real and Fourier space respectively}
\item{{\tt force\_f} stores the force of gravity (short range only) along the three Cartesian directions}
\item{{\tt kern\_f} stores the fine grid force kernel in the three directions}
\item{{\tt PID} stores the unique particle identification tags as double integers.}
\item{{\tt send\_buf\_PID} and {\tt recv\_buf\_PID} store the ID to be {\small MPI}-passed to the neighbouring sub-volumes}
\end{enumerate}
The particle ID is a feature that can be switched off simply by removing a compilation flag, 
and allows to optimize the code for higher resolution or leaner configurations.

In terms of memory, {\small CUBEP$^3$M} can be configured such as the main footprint is dominated by the 
particle {\tt xv} and the linked list arrays ($24$ and $4$ bytes per particles respectively),
with a small contribution needed for the {\tt send\_buf} and  {\tt recv\_buf} arrays (as low as $1$ byte each),
and up $7$ bytes scattered in all the other arrays at peak consumption (i.e. for runs with more than 20 000 processors). 
Enabling particle ID would crank this number by $8$ bytes, for a total of $45$ bytes per particle, assuming no load imbalances.
For a fixed problem size, this extra-lean mode can be achieved by maximizing the number of tiles, thereby reducing the size of the local fine mesh arrays,
and turning off the particle ID and the halo finder. 
The limit at which we can do such a volume breakdown is reached at $96^3$ cells per tile,
since  the physical volume on each tile must be at least twice as large as the buffer around it (see section \ref{sec:Poisson}).

For comparison, Lean-{\small GADGET-2} and Lean-{\small GADGET-3} codes use  $84$ and $72$ bytes per particles respectively 
at peak usage, with particle ID enabled \citep{2005Natur.435..629S, 2012MNRAS.426.2046A}, 
plus a 10-15 per cent overhead for allowing load imbalances along the evolution -- which we also need to factor in. 

%This allows a low rate of particle exchange across nodes;  in the default configuration, the {\tt send\_buf} and  {recv\_buf} arrays 
%are much larger, in which case the memory per particle goes up to $44$ bytes. 
%Enabling the particle ID adds at least 8 bytes per particle from the main array, and as the {\small MPI}-communication buffers are allocated larger sizes,
%they contribute at most an additional 16 bytes as well. Hence a `lean' version with the particle ID tags uses down to $36$ bytes per particles.
In practice, minimizing the memory does not optimize the speed, as many {\small CPU}-intensive operations are performed on each tile -- the {\small FFTW} for instance.
A higher number of cells per tile -- typically $176^3$ - $250^3$ -- with fewer tiles usually runs faster, in which case the other arrays in the above-mentioned list become significantly larger,
and the memory per particle is closer to $90$ to $140$ bytes for a large production run, including the buffers for imbalances and runtime halo finding.

%\subsection{Code overview}
%\label{subsec:overview}

The code flow is presented in Fig. \ref{fig:structure} and \ref{fig:particle_mesh}.
Before entering the main loop, the code starts with an initialization stage, 
in which many declared variables are assigned default values,
the redshift checkpoints are read, the {\small FFTW} plans are created, and the {\small MPI} communicators are defined.
The {\tt xv} array  is obtained from the output of the initial conditions generator (see section \ref{subsec:init}),
and the force kernels on both grids are constructed by reading precomputed kernels that are then adjusted to the specific simulation size.
For clarity, all these operations are collected under the subroutine call {\tt initialize} in Fig. \ref{fig:structure}, 
although they are actually distinct calls in the code.

Each iteration of the main loop starts with the {\tt timestep} subroutine, 
which proceeds to a determination of the redshift jump by comparing the step size constraints from each
force components and from the scale factor, as calculated at the end of the last time step.
As such, the global time stepping is adaptive -- $dt$ evolves with time, although fixed for all particles at a given time step -- and is calculated as $dt \propto (a G a_{max} )^{-1}$, where $a_{max}$
is the maximum acceleration of all force components.
The cosmic expansion is found by Taylor expanding Friedmann's equation up to the third order in the scale factor,
and can accommodate constant or running equation of state of dark energy.
The force of gravity is then solved  in the {\tt particle\_mesh} subroutine,
which first updates the positions and velocities of the dark matter particles, exchange with neighbouring nodes those that have exited to volume,
creates a new linked list, then solve Poisson's equation.  This subroutine is conceptually identical to that of {\small PMFAST}, 
with the exceptions  that {\small CUBEP$^3$M} decomposes the volume into cubes (as opposed to slabs), and computes pp interactions as well. 
The loop over tiles and the particle exchange are thus performed in three dimensions.
When the short range and pp forces have been calculated on all tiles, the code exits the parallel {\small OPENMP} loop
and proceeds to the long range calculation. This section of the code is also parallelized on many occasions, but, unfortunately, the current version
is based on  {\small FFTW-2}, in which the {\small MPI} 
libraries do not allow multi-threading. There is thus an inevitable loss of efficiency during each global Fourier transforms, during which
only the single core {\small MPI} process is active\footnote{The recent development of FFTW-3 now allows for multi-threading operations during the MPI interface,
even though the {\tt transpose} subroutines are not threaded. 
Other libraries such as {\small P3DFFT} ({\tt http://code.google.com/p/p3dfft/}) currently permit
 both the extra level of parallelization, plus a greater scalability, and efforts are currently being invested towards the migration to one of these in the near future.}.
 
 Once the force components have been calculated, the code finds out the largest acceleration, 
 from which it calculates the time step size : large accelerations imply small steps, such that the trajectory of each particle can be well 
 approximated by straight line segments. If the forces are weak everywhere, as it is usually the case early on in a simulation,
 then the step sizes are constrained by a maximum cosmic expansion, $ra_{max}$ (see section \ref{subsec:z_jumps} for details about how this variable is determined).
 In order to resolve adequately the non-linear structures, the particle position and velocity updates are performed in a Runge-Kutta integration scheme,
 which has the advantage to allow for adaptive time stepping compared to the common leap-frog scheme \citep{1981csup.book.....H}, at the cost of loosing time reversibility.

\begin{figure}
\begin{verbatim}
program cubep3m
   call initialize
   do
       call timestep
       call particle_mesh
       if(checkpoint_step) then
          call checkpoint
       elseif(last_step)
          exit
       endif
   enddo
   call finalize
end program cubep3m
\end{verbatim}
\caption{Overall structure of the code, simplified for readability.}
\label{fig:structure}
\end{figure}

\begin{figure}
\begin{verbatim}
subroutine particle_mesh
   call update_position + apply random offset
   call link_list
   call particle_pass
   !$omp parallel do
   do tile = 1, tiles_node
      call rho_f_ngp
      call cmplx_rho_f
      call kernel_multiply_f
      call force_f
      call update_velocity_f
      if(pp = .true.) then       
         call link_list_pp
         call force_pp
         call update_velocity_pp
         if(extended_pp = .true.) then
            call link_list_pp_extended
            call force_pp_extended
            call update_velocity_pp_extended       
         endif
      endif
   end do
   !$omp end parallel do
   call rho_c_ngp
   call cmplx_rho_c
   call kernel_multiply_c
   call force_c
   call update_velocity_c      
   delete_buffers
end subroutine particle_mesh
\end{verbatim}
\caption{Overall structure of the two-level mesh algorithm. We have included the section that concerns the standard pp and the extended pp force calculation, to illustrate that they follow similar linked-list logic. }
\label{fig:particle_mesh}
\end{figure}

If the current redshift corresponds to one of the checkpoints, the code advances all particles to their final location
and writes their positions and velocities on file, preceded by a header that 
contains the local  number of particles, the current redshift and time step, and the constraints on the time step jump
from the previous iteration. The particle identification tags are written 
with a similar fashion in distinct files to simplify the post-processing coding. 
This general I/O strategy allows for highly efficient {\small MPI} parallel read and write
(by default in binary format for compactness) and has been shown to scale well to large data sets. 
At this stage, the performance depends only on the setup and efficiency of the parallel file system.

Similarly, the code can compute two-dimensional projections of the density field, halo catalogues (see section \ref{sec:halo} for details), and can compute the power spectrum on the coarse grid at run time. 
The code exits the main loop when it has reached the final redshift, it then wraps up the {\small FFTW} plans 
and clears the {\small MPI} communicators. We have grouped these operations under the subroutine {\tt finalize} in Fig. \ref{fig:structure} for concision.

Other constraints that need to be considered is that any decomposition geometry 
limits in some ways the permissible grid sizes, and that volumes evenly decomposed into cubic sub-sections 
--  such as {\small CUBEP$^3$M} -- require the number of {\small MPI} processes to be a perfect cube.
In addition, the slab rearrangement that speeds up the global {\small FFTW} requires
the number of coarse cells per slab to be an integer number. For instance, 
a $256^{3}$ cell box can decompose into one slab (slabs are 64 coarse cells thick)
or 8 slabs (slabs are 8 coarse cells thick), but it would not divide into 64 slabs (slabs would be 1/8 coarse cell thick).
We have tabulated in the Appendix \ref{app:geometry} a few examples of internal geometry configurations.
Note that these issues concern only the simulation grid, and the cosmology and volume are free parameters 
to be determined by the user.

Also,  since the decomposition is volumetric, as opposed to density dependent,
it suffers from load imbalance whenever the particles are not evenly distributed.
However, large scale cosmology problems are very weakly affected by this effect, which
can generally  be overcome by allocating a small amount of extra memory per node.
This is done by adjusting the {\tt density\_buffer} parameter, which increases the size
of the particle {\tt xv}, {\tt PID}, {\tt ll}, and {\tt send/recv\_buf} arrays.
As mentioned in section \ref{sec:scaling}, large production runs can also be optimized by  allocating 
less memory to start with, follow by a restart with a configuration that allows more memory per node, if required.

%%%%%%%%%%%%%%%%%%%
%\input{../cubep3m_paper/PoissonSolver}

\section{Poisson Solver}
\label{sec:Poisson}

This section reviews how Poisson's equation is solved on a double-mesh configuration. 
Many parts of the algorithm are identical to {\small PMFAST}, hence we refer the reader 
to section 2 of MPT for more details. In {\small CUBEP$^3$M}, the mass default assignment scheme are
a `cloud-in-cell' (CIC) interpolation for the coarse grid,  and a `nearest-grid-point' (NGP) interpolation 
for the fine grid \citep{1981csup.book.....H}. This choice is motivated by the fact that the most straightforward 
way to implement a P$^3$M algorithm on a mesh is to have exactly zero mesh force inside a grid, 
which is only true for the NGP interpolation. Although CIC generally has a smoother and more accurate force,
the pp implementation enhances the code resolution by almost an order of magnitude. 

The code units inherit from \citep{2004NewA....9..443T} and are summarized here for completeness.
The comoving length of a fine grid cell is set to one,
such that the unit length in simulation unit is 
\begin{eqnarray}
1\mathcal{L} = a \frac{L}{N} 
\end{eqnarray}
where $a$ is the scale factor, $N$ is the total number of cells along one dimension,
and $L$ is the comoving volume in $h^{-1}\mbox{Mpc}$.
The mean comoving mass density is also set to unity in simulation units, 
which, in physical units, corresponds to 
\begin{eqnarray}
1\mathcal{D} = \rho_{m}(0) a^{-3} = \Omega_{m} \rho_{c} a^{-3} = \frac{3 \Omega_{m} H_{o}^{2}}{8 \pi G a^{3} }
\end{eqnarray}
$\Omega_{m}$ is the matter density, $H_{o}$ is the Hubble's constant, $\rho_{c}$ is the critical density today,
and $G$ is Newton's constant. The mass unit is found with $\mathcal{M} = \mathcal{DL}^{3}$.
Specifying the value of $G$ on the grid fixes the time unit, and with $G_{grid}$ = 1/$(6 \pi a)$,
we get:
\begin{eqnarray}
1 \mathcal{T} = \frac{2a^{2}}{3}\frac{1}{\sqrt{\Omega_{m}H_{o}^{2}}}
\end{eqnarray}
These choices completely determine the convertion between physical and simulation units.
For instance, the velocity units are given by $1\mathcal{V} = \mathcal{L}$/$\mathcal{T}$.

The force of gravity on a mesh can be computed either with a gravitational potential  kernel $\omega_{\phi} ({\bf x})$
  or a force  kernel $\omega_{F} ({\bf x})$.
Gravity fields are curl-free, which allows us to relate the potential $\phi({\bf x})$ to the source term via Poisson's equation: 
\begin{eqnarray}
\nabla^{2}\phi({\bf x}) = 4 \pi G \rho({\bf x})
\label{eq:poisson}
\end{eqnarray}
We solve this equation in Fourier space, where we write:
\begin{eqnarray}
 \tilde{\phi}({\bf k}) = \frac{-4 \pi G \tilde{\rho}({\bf k})}{k^{2}} \equiv \tilde{\omega}_{\phi}({\bf k})\tilde{\rho}({\bf k})
\label{eq:poissonFourier}
\end{eqnarray}
The potential in real space is then obtained with an inverse Fourier transform, and the kernel becomes $\omega_{\phi} ({\bf x}) = -G/r$,
with $r = |{\bf x}|$.
Using the convolution theorem, we can write
\begin{eqnarray}
 \phi({\bf x}) = \int \rho({\bf x'}) \omega_{\phi}({\bf x'} - {\bf x}) d{\bf x'}   
\label{eq:poisson_solution_pot}
\end{eqnarray}
and
\begin{eqnarray}
{\bf F}({\bf x}) = - m {\bf \nabla} \phi({\bf x}) 
\label{eq:Force_sol}
\end{eqnarray}
Although this approach is fast, it involves a finite differentiation at the final step, which enhances the numerical noise.
We therefore opt for a force kernel, which is more accurate, even though it requires four extra Fourier transforms.
In this case, we must solve the convolution in three dimensions and define the force kernel {\boldmath $\omega$}$_{F}$ such as:
\begin{eqnarray}
 {\bf F}({\bf x}) =  \int \rho({\bf x'}) \mbox{\boldmath $\omega$}_{F}({\bf x'} - {\bf x}) d{\bf x'}                                      
\label{eq:poisson_solution_force}
\end{eqnarray}
Because the gradient acting on Eq. \ref{eq:poisson_solution_pot} affects only unprime variables, we can express the force kernel as a gradient of the potential kernel. Namely:  
\begin{eqnarray}
\mbox{\boldmath $\omega$}_{F}({\bf x}) \equiv - {\bf \nabla}\omega_{\phi}({\bf x}) = - \frac{mG \hat{\bf r}}{r^{2}}
\end{eqnarray}

Following the spherically symmetric matching technique of MPT (section 2.1), 
we split  the force kernel into two components, for the short and long range respectively, and 
match the overlapping region with a polynomial. Namely, we have:
\begin{eqnarray}
\mbox{\boldmath $\omega$}_{s}(r) = \begin{cases} \mbox{\boldmath $\omega$}_{F}(r) -  \mbox{\boldmath $\beta$}(r) &\mbox{if  } \mbox{$r$ $\le$ $r_{c}$ } \\
0 & \mbox{otherwise} 
\end{cases}
\end{eqnarray}
and
\begin{eqnarray}
\mbox{\boldmath $\omega$}_{l}(r) = \begin{cases} \mbox{\boldmath $\beta$}(r) &\mbox{if  } \mbox{$r$ $\le$ $r_{c}$ } \\
 \mbox{\boldmath $\omega$}_{F}(r)  &\mbox{otherwise} 
\end{cases}
\end{eqnarray}
The vector $\mbox{\boldmath $\beta$}(r)$ is related to the fourth order polynomial that is used in the potential case described in MPT by
 $ \mbox{\boldmath $\beta$} = - {\bf \nabla} \alpha(r)$. The coefficients are found by matching the boundary conditions at $r_{c}$ up to the second derivative,
 and we get
  \begin{eqnarray}
   \mbox{\boldmath $\beta$}(r) = \bigg[ -\frac{7 r}{4 r_{c}^{3}} + \frac{3 r^{3}}{4 r_{c}^{5}}\bigg] \hat{\bf r}
  \end{eqnarray}

Since these calculations are performed on two grids of different resolution, a sampling window function must be convoluted 
both with the density and the kernel (see [Eq. 7-8] of MPT).
When matching the two force kernels, the long range force is always on the low side close to the cutoff region, whereas the short range force is uniformly scattered across the theoretical $1/r^2$ value -- intrinsic features of the CIC and NGP interpolation schemes respectively.  By performing force measurements on two particles randomly placed in the volume, we identified a small region surrounding the cutoff length in which we empirically adjust both kernels such as to improve the match. Namely, for $14 \le r \le 16$ in fine cell units, $\mbox{\boldmath $\omega$}_{s}({r}) \rightarrow 0.985\mbox{\boldmath $\omega$}_{s}({ r})$,
and for  $12 \le r \le 16$, $\mbox{\boldmath $\omega$}_{l}({r}) \rightarrow 1.2\mbox{\boldmath $\omega$}_{l}({ r})$.
A recent adjustment of the similar kind was also applied on the fine kernel to compensate for a systematic underestimate of the force at fine grid distances.
This is caused by the uneven balance of the force about the $1/r^2$ law in the NGP interpolation scheme.
Although most figures in this paper were made prior this correction, the new code configuration applies a $60$ per cent boost to 
the kernel for the elements that are up to 2 layers adjacent, i.e. $\mbox{\boldmath $\omega$}_{s}({r\le2}) \rightarrow 1.6\mbox{\boldmath $\omega$}_{s}({r\le2})$ (see section \ref{subsec:force} for more details).
As mentioned in section \ref{sec:structure} and summarized in Fig. \ref{fig:particle_mesh},
the force kernels are first read from files in the code initialization stage.
Eq. \ref{eq:poisson_solution_force} is then solved with fast Fourier transforms along each direction, 
 and is applied onto particles in the {\tt update\_velocity} subroutine.

The pp force is calculated during the fine mesh velocity update, which avoids loading the particle list twice and allows the operation to be threaded without significant additional work. During this process, the particles within a given fine mesh tile are first read in via the linked list, 
then their velocity is updated with the fine mesh force component, according to their location within the tile. 
In order to organize the particle-particle interactions, we proceed by constructing a set of threaded fine-cell linked list chains for each coarse cell. 
We then calculate the pairwise force between all particles that lie within the same fine mesh cell, excluding pairs whose separation is smaller than a softening length $r_{soft}$; particles separated by less than this distance thus have their 
particle-particle force set to zero.  As this proceeds, we accumulate the  pp force applied on each particle and then determine the maximum force element of the pp contribution, which is also taken into account when constraining the length of the global time step. 

\subsection{Force Softening}
\label{subsec:force_soft}

Force softening is generally required by any code to prevent
large scattering as $r \rightarrow 0$ that can otherwise slow the calculation down, 
and to reduce the two-body relaxation, which can affect the numerical convergence. 
Many force softening schemes can be found in the literature, including Plummer force, uniform or linear density profiles or the spline-softened model 
(see \citet{1993ApJ...409...60D} for a review). In the current case, a sharp force cutoff corresponds to a particle interacting with a hollow shell.
In comparison with other techniques, this  force softening is the easiest to code and the fastest to execute. 
Generally, it is desirable to match the smoothness of the force to the order of the time integration. 
A Plummer force is infinitely differentiable, which is a sufficient but not necessary condition for our $2^{nd}$ order time integration.  
Also, one of the drawbacks of Plummer's softening is that the resolution degrades smoothly: the effects of the smoothing are present at all radii. 
In comparison, the uniform density and hollow shell alternatives both have the advantage that the deviations from $1/r^2$ are minimized. 
Although all the results presented in this paper were obtained with the sharp cutoff softening, other schemes can easily be adopted as these 
are typically single line changes to the code. \citet{1993ApJ...409...60D} argues in favour of the uniform density profile scheme -- which is more physical --
and future developments of {\small CUBEP$^3$M} will incorporate this option.

\begin{figure}%[ht]
  \begin{center}
    \includegraphics[width=3.2in]{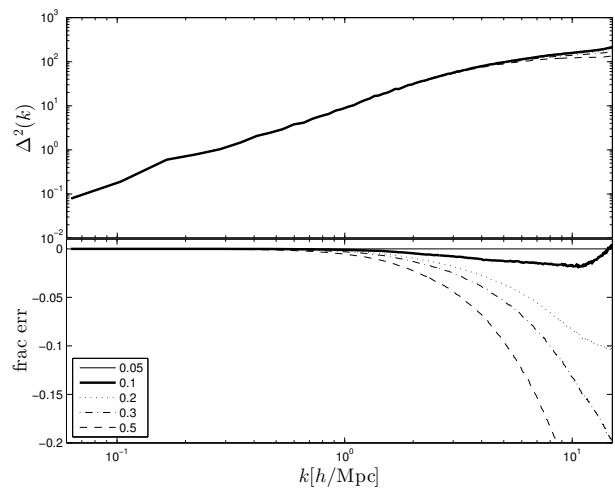}
  \caption{({\it top}:) Dark matter power spectrum, measured at $z=0.5$ in a series of SCINET256 simulations in which the softening length is varied.
  The simulations all  started off at $z_i= 100$ with the same initial conditions, and have a box size of $100 h^{-1}\mbox{Mpc}$.
  ({\it bottom}:) Fractional error with respect to the $r_{soft} = 0.05$ simulation;  the default value is represented by the thick solid line.
    \label{fig:rsoft}}
\end{center}
\end{figure}

The choice of softening length is motivated by a trade off between accuracy and run time.
 Larger values reduce the structure formation but make the code run quicker. 
 We show in Fig. \ref{fig:rsoft} the impact of changing this parameter on the power spectrum.
 For this test, we produced a series of SCINET256 simulations, each starting at $z=100$ and evolving to $z=0.5$, with a box size of $100 h^{-1}\mbox{Mpc}$.
 They each read the same set of initial conditions and used the same random seeds (see section \ref{subsec:force} for more details), such that the only difference between them is the softening length.
 We record in Table \ref{table:rsoft} the real time for each trials, from where 
we witness the strong effect  of the choice of $r_{soft}$ on the computation time.
In this test, which is purposefully probing rather deep in the non-linear regime, reducing the softening length
to half its default value of $1/10^{th}$ of a fine grid cell doubles the run time, while the effect on the power spectrum is less than two per cent.
Similarly, increasing $r_{soft}$ to $0.2$ reduces the run time almost by a half, but suffers from a five to ten per cent loss in power at small scales.
As expected, this resolution loss gets worst if the softening length is made larger.
One tenth of grid cell seems to be the optimal choice in this trade off, however it should really be considered a free parameter.

\begin{table}
\begin{center}
\caption{Scaling in {\small CPU} resources as a function of the softening length, in units of fine grid cells and in units of mean inter-particle separation $\langle r \rangle = L/N_{particles}$.}
\begin{tabular}{|l|c|c|}
\hline 
$r_{soft}$    & $r_{soft} / \langle r \rangle$  & time (h)   \\                 
\hline
 $0.5$ & 0.250 & 1.71 \\
 $0.3$ & 0.150 & 2.28\\
 $0.2$ & 0.100 & 2.96 \\
 $0.1$ & 0.050 & 4.93  \\
 $0.05$ & 0.025& 8.14 \\
% $0.03$ & XXX\\
% $0.01$ & XXX\\
\hline
\end{tabular}
\label{table:rsoft}
\end{center}
\end{table}

In addition, {\small PMFAST} could run with a different set of force kernels, described in MPT as  `least square matching', 
a technique which adjusts the kernels on a cell-by-cell basis such as to minimize the deviation with respect
to Newtonian predictions. This was originally computed in the case where both grids are obtained from CIC interpolation. Moving to a mix CIC/NGP scheme
requires solving the system of equations with the new configuration, a straightforward operation.
 With the inclusion of the random shifting (see section \ref{subsec:force}), however, it is not clear how much improvement 
one would recover from this other kernel matching.  It is certainly something
we will investigate and document in the near future.

Finally, a choice must be done concerning the longest range of the coarse mesh force. Gravity can be either a) a $1/r^2$ force as far as the volume allows, 
or b) modified to correctly match the periodicity of the boundary conditions. By default, the code is configured along to the second choice,
which accurately models the growth of structures at very large scales. However, detailed studies of gravitational collapse of a single large object would benefit 
from the first setting, even though the code is not meant to evolve such systems as they generally require load balance control.

%%%%%%%%%%%%%%%%%%%%%%%%
%\input{../cubep3m_paper/ScalingPerformance}

\section{Scaling Performances}
\label{sec:scaling}

\begin{figure*}%[ht]
%  \vskip -0.5cm 
  \begin{center}
    \includegraphics[width=3.0in]{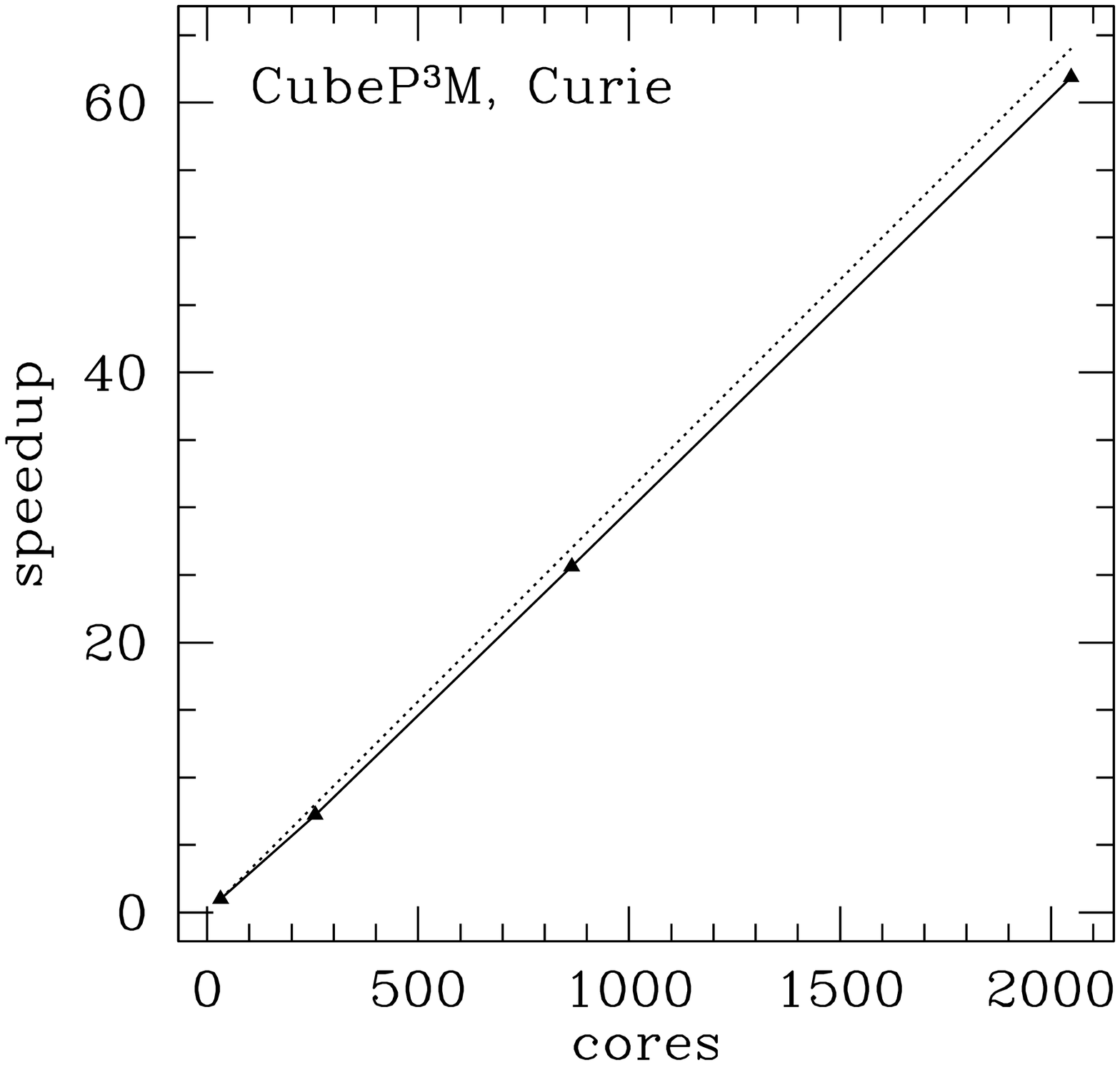}
    \includegraphics[width=3.0in]{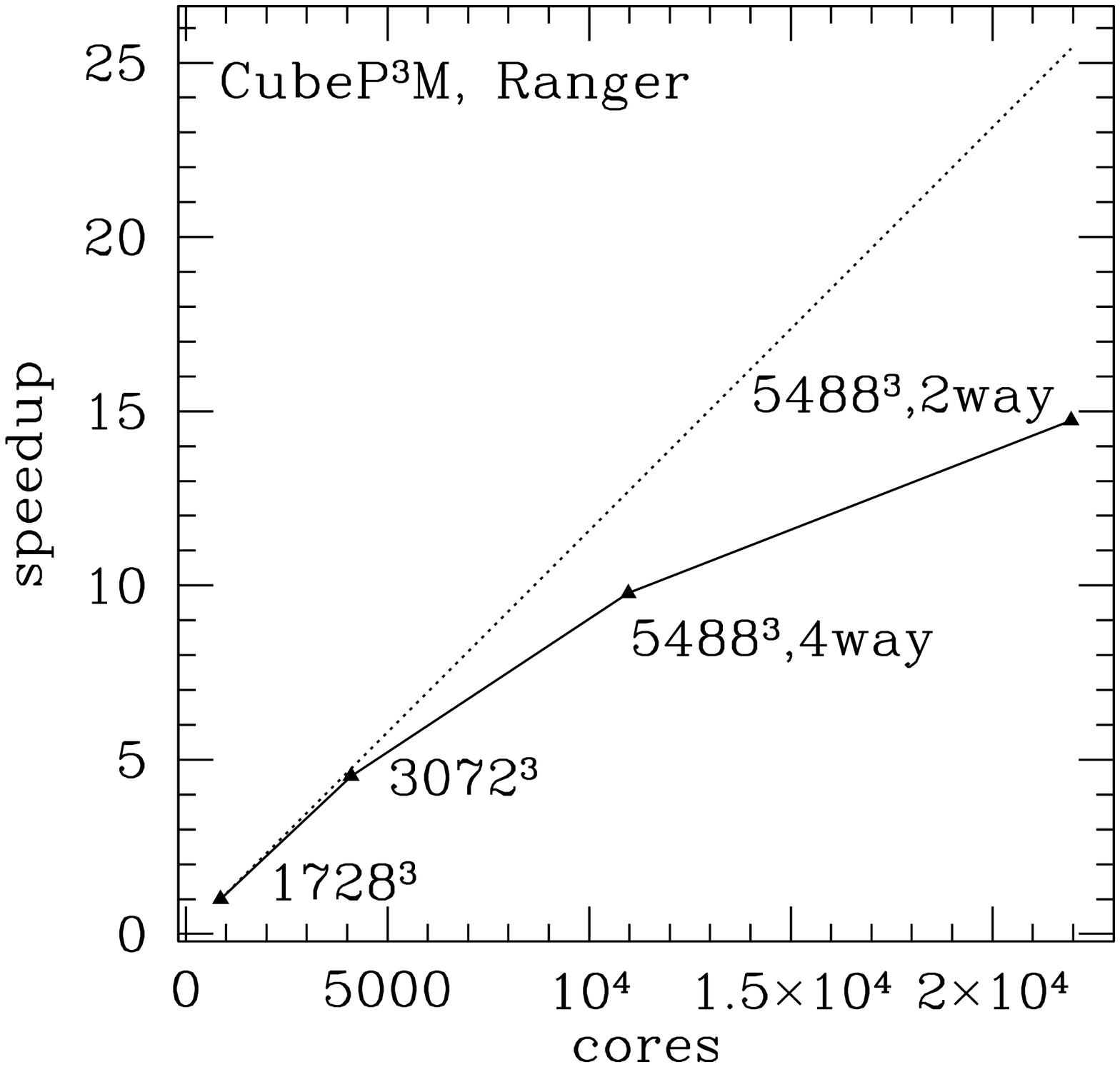}
%  \vskip -1.2cm 
%  \vskip -0.5cm 
  \caption{Scaling of {\small CUBEP$^3$M} on Curie fat nodes (left) and 
    on Ranger TACC facility for very large number of cores (right). Plotted is the code speedup 
    ($N_{\rm particles}^3/t_{\rm wallclock}$) against core count, normalized by the smallest run 
    in each case. The dashed line indicates the ideal weak 
    scaling, and details on the data are listed in Table \ref{summary_scaling_table}.
    \label{scaling}
% \vskip -0.9cm 
}
\end{center}
\end{figure*}

\begin{table*}%[ht]
  \vskip -0.5cm 
  \begin{center}
\caption{Scaling of {\small CUBEP$^3$M} on Curie. Speedup is 
scaled to the smallest run.}
\label{summary_scaling_table}
\begin{tabular}{@{}|llllll|}
\hline
number of cores & speedup & ideal speedup & absolute timing (min) & 
$N_{\rm particles}$& box size ($h^{-1}$Mpc)
\\[2mm]\hline
%\hline
32  &  1.00 & - &3.2 & $256^3$ & 256\\
256  & 7.21 & 8 &3.55 & $512^3$  & 512\\
864  & 25.63 & 27 &4.8 & $864^3$  & 864\\
2048  & 61.87 & 64 &26.48 & $2048^3$ & 2048 \\
\hline
\end{tabular}
\caption{Scaling of  {\small CUBEP$^3$M} on Ranger. Speedup is scaled to the smallest run.}
\label{summary_scaling_table2}
\begin{tabular}{@{}|llllll|}
\hline
number of cores & speedup & ideal speedup & absolute timing (min) & 
$N_{\rm particles}$& box size ($h^{-1}$Mpc)
\\[2mm]\hline
%\hline
864    & 1.00  & -    &258   & $1728^3$  & 6.3\\
4096   & 4.53  & 4.74 &320   & $3072^3$  & 11.4\\
10976  & 9.78  & 12.7 &845   & $5488^3$  & 20\\
21952  & 14.73 & 25.4 &561   & $5488^3$  & 20 \\
\hline
\end{tabular}
\end{center}
  \vskip -0.7cm 
\end{table*}

%the Ranger system at the Texas Supercomputing
%Center (in top 20 in the world) which is a SunBlade x6420 
%with AMD x86\_64 Opteron Quad Core 2300 MHz (9.2 GFlops)
% ``Barcelona'' processors and Infiniband networking. It has 
%a total of 62976 computing cores and 125952 GB of total 
%memory. Its nodes consist of 4 Quad Core processors and 32 GB 
%of shared RAM. For efficiency reasons (local memory access) we 
%typically use smaller MPI 'nodes' consisting of one Quad Core 
%processor and 8 GB of RAM. 

The parallel algorithm of {\small CUBEP$^3$M} is designed for `weak' 
scaling, i.e. if the number of cores and the problem size 
increase in proportion to each other, then for ideal scaling the 
wall-clock time should remain the same. This is to be in contrasted with `strong' 
scaling codes, whereby the same problem solved on more cores should take 
proportionally less wall-clock time. This weak scaling requirement 
is dictated by the problems we are typically investigating (very 
large and computationally-intensive) and our goals, which are to 
address such large problems in the most efficient way, rather than 
for the smallest wall-clock time. Furthermore, we recall that there is no explicit 
load balancing feature, thus the code is maximally efficient when the sub-domains
contain roughly an equal number of particles. This is true for most
cosmological size volumes that do not resolve too deep in the non-linear regime, 
but not for e.g. simulations of a single highly-resolved galaxy. 

We first recall that because of the volumetric decomposition, the total number of {\small MPI} processes needs
to be a perfect cube. Also, for maximal resource usage, the number of tiles per node 
should be a multiple of the number of available {\small CPU}s per {\small MPI} process,
such that no core sits idle in the threaded block.
Given the available freedom in the parallel configuration, as long as the load is balanced, 
it is generally good practice to maximize the number of {\small OPENMP} threads and minimize the number of {\small MPI} processes:
this reduces the total amount of buffer zones, freeing memory that can be used to increase the mesh resolution. 
In addition, the particles passing across nodes are done with larger but fewer arrays, which 
optimizes the communication cost in most systems.
However, it has been observed that for the case of non-uniform memory access (NUMA) systems, 
the code is optimized when only the cores that share the same socket are  {\small OPENMP} threaded.
As one probes deeper into the non-linear regime however, 
the formation of dense objects can cause memory problems in such configurations, and increasing 
the number of {\small MPI} processes helps to ensure memory locality,
especially in NUMA environments.

%Since the current Curie is using Intel Nehalem architecture, while 
%the thin Curie nodes will use the newer Westermere Intel architecture, 
%we also show the scaling of our code on the Westermere-based Lonestar 
%computer at the Texas Advanced Computing Centre 

 The intermediary version of the code -- {\small CUBEPM} --
was  ported to the IBM Blue Gene/L platform, and achieved 
weak-scaling up to 4096 processes (over a billion particles), with the N-body calculation incurring only a 10 per cent overhead 
at runtime (compared to 8 processes) for a balanced workload\footnote{\tt http://web.archive.org/web/20060925132146/ http://www-03.ibm.com/servers/deepcomputing/pdf/ Blue\_Gene\_Applications\_Paper\_CubePM\_032306.pdf}.  In order to 
accommodate the limited amount of memory available per processing core on the 
Blue Gene/L platform machines, it was necessary to perform the long range {\small MPI FFT}
with a volumetric decomposition \citep{3DFFT}.
Slab decomposition would have required a volume too large to fit in system 
memory given the constraints in the simulation geometry.

The scaling of {\small CUBEP$^3$M}  was first  tested with a dedicated series 
of simulations -- the CURIE simulation suite-- by increasing the size and number of cores on the `fat' 
(i.e. large-memory) nodes of the  Curie supercomputer at the Tr\`{e}s Grand Centre de Calcul (TGCC) in France. 
 For appropriate direct comparison, all these simulations were performed using the same particle mass 
($M_{\rm particle}=1.07\times10^{11}M_\odot$) and force resolution (softening length of 50 comoving $h^{-1}$kpc). 
This fixes the amount of non-linear structure that forms in each local volume between trials, 
and which generally becomes the main bottle neck in the evolution.
The box sizes ranged from 256 $h^{-1}$Mpc
to 2048 $h^{-1}$Mpc, and the corresponding number of particles from $256^3$ to $2048^3$.
Simulations were run on 32 up to 2048 computing cores, all starting from 
redshift $z=100$, and evolving until $z=0$. 
Under that setup, the number of time steps across the tests
is close to being a constant, such that the timing differences actually come 
from the scaling performances.
Our results are shown in Fig. \ref{scaling} and in Table \ref{summary_scaling_table}, and present excellent scaling, within 
$\sim3$ per cent of ideal, at least for up to 2048 cores.

We have also ran {\small CUBEP$^3$M} on a much larger number of cores, 
from 8000 to up to 21,976, with $5488^3$-$6000^3$ (165 to 216 billion) 
particles on Ranger and on JUROPA at the J\"ulich Supercomputing Centre in Germany, 
an Intel Xeon X5570 
(Nehalem-EP) quad-core 2.93 GHz system, also interconnected with Infiniband.
Since it is not practical to perform dedicated scaling tests on such a large number of
computing cores, we instead list in Table \ref{summary_scaling_table2} 
the data directly extracted from production runs. We have found the 
code to scale within 1.5 per cent of ideal up to 4096 cores. 
For larger sizes ($\ge$10,976 cores), the scaling is less ideal,
due to increased communication 
costs, I/O overheads (a single time slice of $5488^3$ particles is 3.6 TB)
and load balancing issues, but still within $\sim20$ per cent of ideal. 
These first three Ranger runs were performed 
with 4 {\small MPI} processes and 4 threads per Ranger node (`4way')\footnote{For these very large runs, 
we used a NUMA script {\it tacc\_affinity}, specially provided by the technical staff, 
that bind the memory usage to local sockets, thus ensuring memory affinity. 
This becomes important because the memory sockets per node 
(32 GB RAM/node on Ranger) are actually not equal-access. Generally, the local 
memory of each processor has much shorter access time.}.

Furthermore, due to the increasing clustering of structures at those small 
scales, some of the cuboid sub-domains came to contain a number of particles
well above the 
average, thereby requiring more memory per {\small MPI} process
in order to run until the end. 
As a consequence,  throughout most of their late evolution, the 
largest two of these simulations were run with 4096 and 21,952 cores and 
with only 2 {\small MPI} processes and 8 threads per node (`2way'), which on 
Ranger allows using up to 16 GB of RAM per {\small MPI} process\footnote{In order to ensure 
local memory affinity,  a second special NUMA control script, {\it tacc\_affinity\_2way}, 
was developed for us by the TACC technical staff and allowed to run more efficiently 
in this mode.}. Because each processor accesses memory that is not fully local, this configuration does affect the 
performance somewhat, as does the imperfect load balancing that arises in such situations.
This can be seen in the rightmost point of Fig. \ref{scaling} (right panel), where the scaling is $42$ per cent below ideal.
We note that we still get $\sim1.5$ speedup
from doubling the core count, even given these issues. 

The code has recently been running even larger core counts.
The JUBILEE6000 run evolved $6000^3$ particles on $1000$ nodes with 24GB RAM each, at the
Neumann Institute for Computing in J$\ddot {\mbox {u}}$lich  as 
detailed in \citet{2012arXiv1212.0095W}.
The BGQ6912 test run was executed on the BlueGene/Q supercomputer located at SciNet,
which consists of compute nodes that each have 16 x 1.6GHz PowerPC based CPU (PowerPC A2) with 16GB of RAM.
{\small CUBEP$^3$M} used a total of $27 648$ out of the $32 768$ available cores, evolving $6912^3$ particles. 
These two runs were dumping about $4$ and $8$ TB per particle dump respectively. 
As mentioned earlier, doing scaling tests at this level is rather expensive,
hence we only have a single production run for these two large scale simulations, 
which can therefore not be added in a meaningful manner to the charts of Table \ref{summary_scaling_table} or  \ref{summary_scaling_table2}. 

Overall, the code scaling performance is thus satisfactory, although not ideal, even at extremely large number of cores.
 We expect the code to handle even larger problems efficiently, and seems well suited to run on
  next generation Petascale systems. 
Finally, we note that several special fixes had to be developed by  the TACC,  JUROPA and SciNet 
technical staff in order for our largest runs to work properly.
In particular, we encountered unexpected problems from software libraries such as 
{\small MPICH} and {\small FFTW} when applied to calculations of such 
unprecedented size. 

%%%%%%%%%%%%%%%%%%%%%%%
%\input{../cubep3m_paper/Systematics}

\section{Accuracy and Systematics}
\label{sec:accuracy}
 
 This section describes the systematics effects that are inherent to our P$^{3}$M algorithm.
 We start with a demonstration of the accuracy with a series of power spectrum measurements, extracted from two large scale production runs,
  and a series of medium size simulations. 
 The halo mass function also assess the code capacity to model gravitational collapse, 
 but depends on the particular halo finder used.
 We thus postpone the discussion on this aspect until section \ref{sec:halo},
and focus for the moment on the particles only. 

In the second part of this section, we quantify the accuracy of the force calculation
 with comparisons to Newton's law of gravitation.
 Since most of the systematics come from the contribution at smaller distances,
 we dedicate a large part of this section to the force calculation at the grid scale. 
 We finally explain how the correction on the fine force kernel at $r\sim1$ was obtained,
 and present how constraining the size of the redshift jumps can help improve the accuracy of the code.
 
 \subsection{Density power spectrum}
 \label{subsec:powerspectrum}
 
One of the most reliable ways to assess the simulation's accuracy at evolving particles
is to measure the density power spectrum at late time, and compare against non-linear predictions. 
For an over-density field $\delta({\bf  x})$, the power spectrum is extracted from the two point function in Fourier space as:
\begin{eqnarray}
\langle | \delta ({\bf k}) \delta ({\bf k'}) | \rangle = (2\pi)^{3}P({\bf k})\delta^3_{D}({\bf  k'} - {\bf k})
\label{eq:power}
\end{eqnarray}
where the angle bracket corresponds to an ensemble (or volume) average.
When computed on a grid, the power spectrum is strongly affect by the mass assignment scheme and it is important
to undo this effect. As proposed by \citet{1981csup.book.....H}, this can be done by dividing the measured power spectrum
by an appropriate kernel. The {\small CUBEP$^3$M} code package comes with a
post-processing power spectrum utility, plus an optional  on-the-fly coarse mesh power spectrum calculator which operates at each time step.
In both cases, the particles are assigned on the density field with a CIC scheme, 
hence the power spectrum is computed as 
\begin{eqnarray}
P({\bf k}) = \frac{|\delta({\bf k})|^2}{H^4(k_x)H^4(k_y)H^4(k_z)} , H(x) = \mbox{sinc}\bigg(\frac{\pi x}{nc}\bigg)
\end{eqnarray}

\begin{figure*}%[ht]
  \begin{center}
    \includegraphics[width=5.2in]{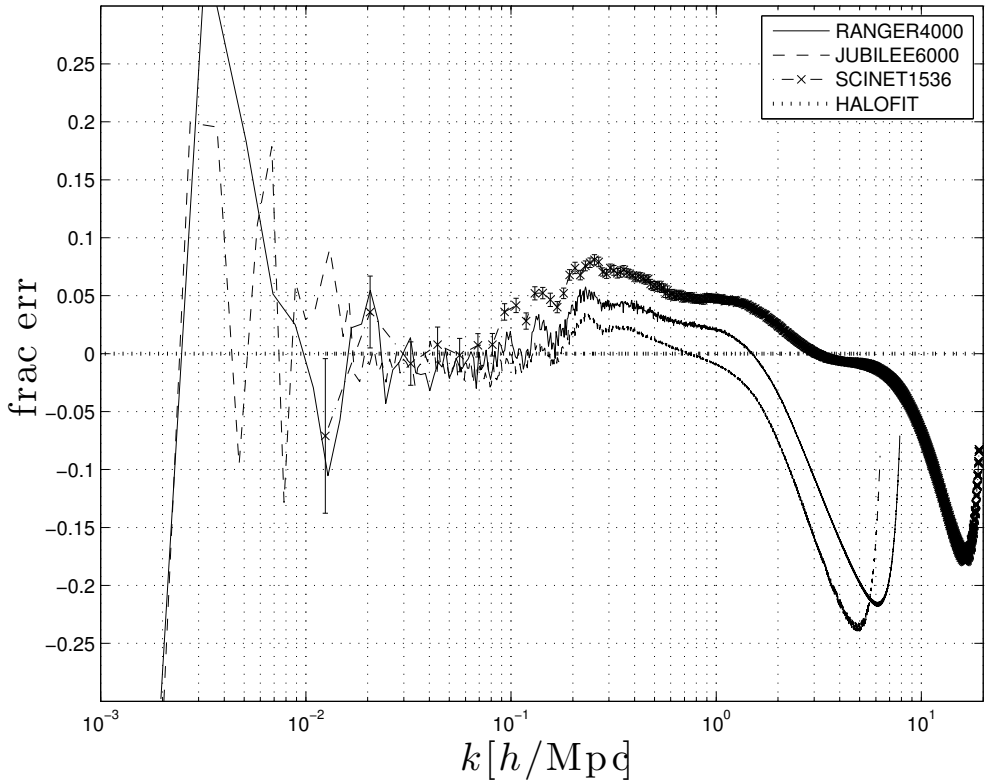}
  \caption{%({\it left}:) 
  Dark matter power spectrum, measured at $z=0$ in the RANGER4000 and the JUBILEE6000 simulations,
  and in an ensemble of 54 SCINET1536 simulations at $z = 0.042$. The error bars are the error on the mean, i.e. $\sigma / \sqrt{54}$.
  Results are presented in the form of  fractional error with respect to the non-linear predictions of {\small HALOFIT} (calculated with the online {\small CAMB} toolkit). 
  %The vertical line corresponds to the scale of the coarse mesh, while the dotted line represents the Poisson noise (not subtracted).
  The particle masses are of $5.7 \times 10^{10}$,  $10.7 \times 10^{10}$ and $3.94 \times 10^9 M_{\odot}$ respectively.
  Poisson noise has not been subtracted, since it is orders of magnitude lower than the signal over the scales that are well resolved by the simulation. 
    \label{fig:power_highres}}
\end{center}
\end{figure*}

Fig. \ref{fig:power_highres} presents  the dimensionless power spectrum, defined as $\Delta^{2} (k) = k^{3}P(k)/(2\pi^{2})$, of 
three distinct setups. First, the RANGER4000 simulation, and JUBILEE6000 simulation,
 which evolved $4000^3$ and $6000^3$ particles respectively until redshift zero.
We observe that the agreement with the non-linear prediction of \citet[{\small HALOFIT} hereafter]{2003MNRAS.341.1311S} is at the 1-2 per cent level in the range $ 0.03 < k = 0.2 h \mbox{Mpc}^{-1}$, 
and within five per cent up to $k = 1.0 h \mbox{Mpc}^{-1}$, after which the power drops because of the finite mesh resolution. 
Note that the Poisson noise has not been subtracted, since it is orders of magnitude lower than the signal.
It is difficult to assess  from a single simulation the accuracy of modes associated with the largest scales,
as these are subject to a large sample variance that comes from the white noise imposed in the initial conditions.
We thus plot the power spectrum measured from an ensemble of 54 SCINET1536 simulations,
which evolved $1536^3$ particles from $z=100$ to $z=0.042$. We observe an excellent agreement between the mean of the measurements
and {\small HALOFIT} in the linear regime, and a slight over-estimate by up to eight per cent in the range $0.1 < k < 2.0 h \mbox{Mpc}^{-1}$.
The same conclusions were drawn from simulations with different resolutions and box sizes \citep{2012MNRAS.419.2949N, 2012MNRAS.423.2288H, 2012MNRAS.426.1262H},
which have shown that samples of a few hundreds of realizations average to within a few per cent of correct value, provided the initial redshift is not too high. 

The few per cent over estimate measured in the range $ 0.1 < k < 1.0 h\mbox{Mpc}^{-1}$ was also observed in the three references above mentioned.
We have checked that this was not a systematic effect caused by inaccuracies at the early time steps,
 as starting the simulation at a later redshift, or running in more accurate modes -- by requiring smaller redshift step size of increasing the range of the pp force --
 do not resolve this discrepancy. It was found independently with other N-body codes that {\small HALOFIT} seems to underestimate by about 5 per cent the 
 power spectrum for $k>0.1 h \mbox{Mpc}^{-1}$ \citep{2007ApJ...665..887M, 2010ApJ...715..104H, 2012ApJ...761..152T}, in good agreement with our measurements.
 Based on these results, we therefore conclude that {\small CUBEP$^3$M} is a cosmological tool that allows for few (i.e. $<5$) per cent precision calculations that rely on power spectrum measurements.

%\begin{figure*}%[ht]
 % \begin{center}
%\includegraphics[width=5.2in]{./fig5.eps}
 % \caption{Detail of a dark matter density $3.2 h^{-1}\mbox{Gpc}$Ê per side, projected over $15 h^{-1}\mbox{Mpc}$ and measured at $z=0$ in a RANGER4000 simulation.
  % In the online version, particles are shown in pink, haloes in blue, and the colour scale is linear. The full projection is 64 times larger,
 % but we show here only a sub-volume, which provides more details on the structure.
   % \label{fig:density}}
%\end{center}
%\end{figure*}

\subsection{Mesh force at grid distances}
\label{subsec:force}

\begin{figure*}%[ht]
  \begin{center}
    \includegraphics[width=3.0in]{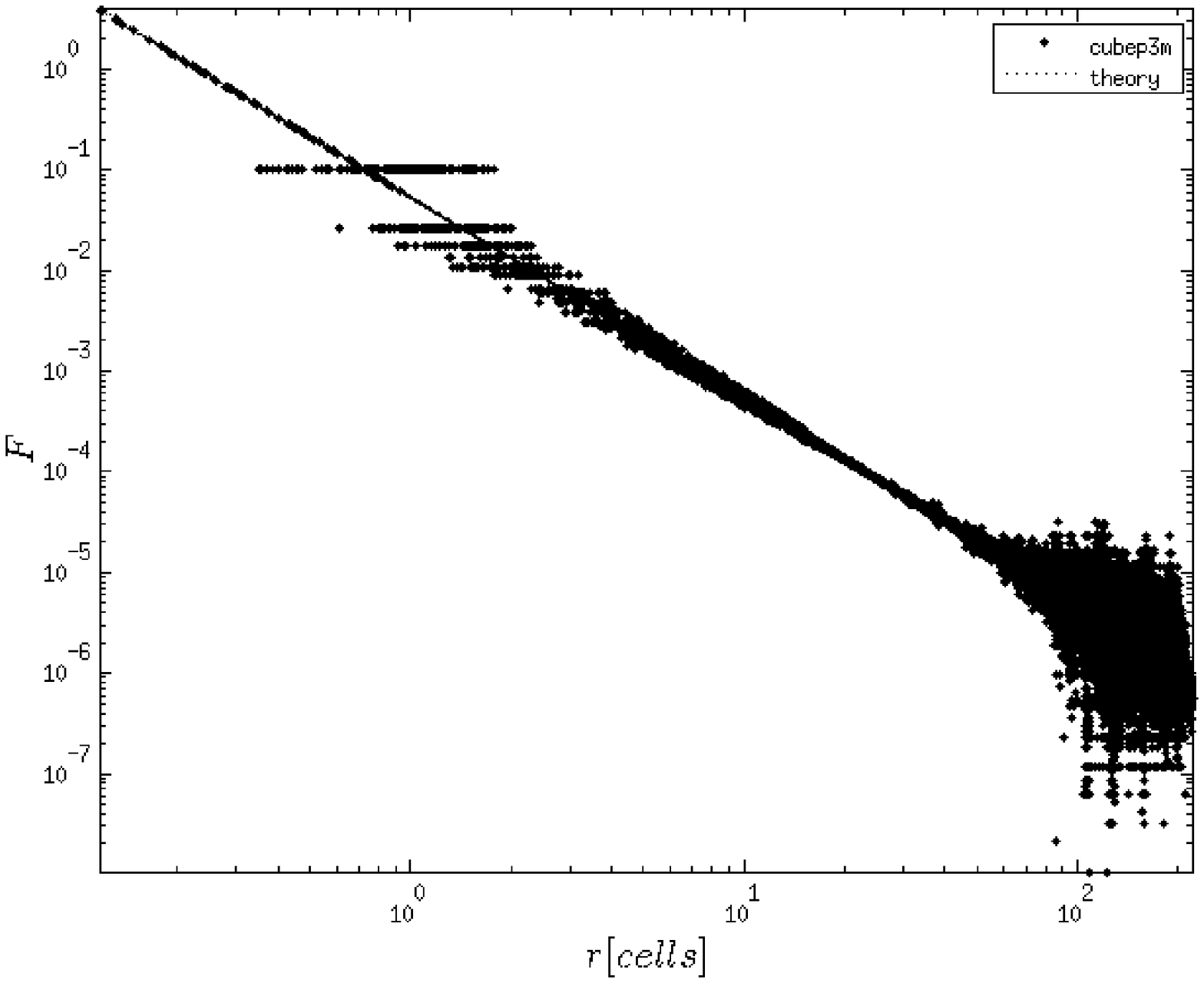}
    \includegraphics[width=3.0in]{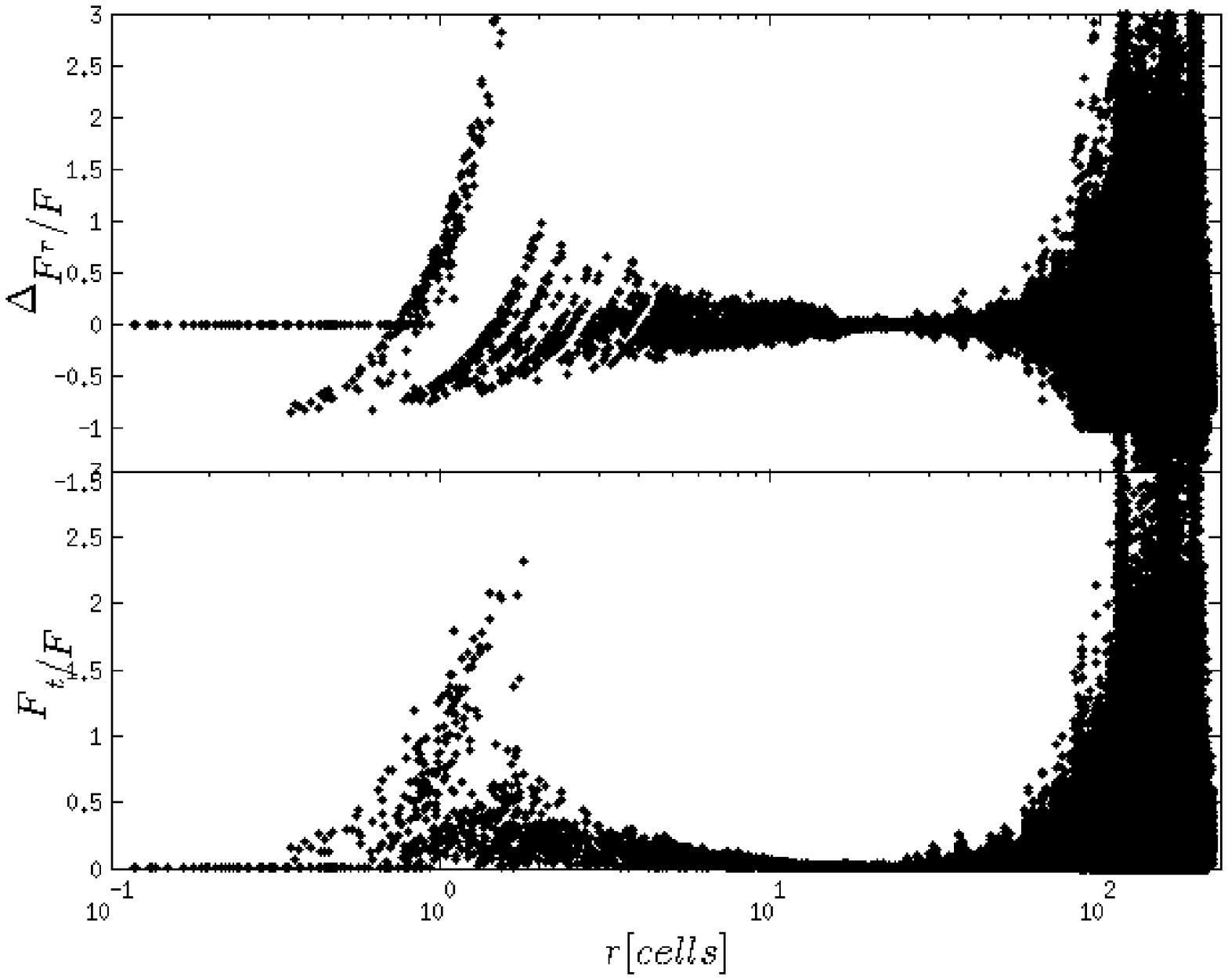}
   \includegraphics[width=3.0in]{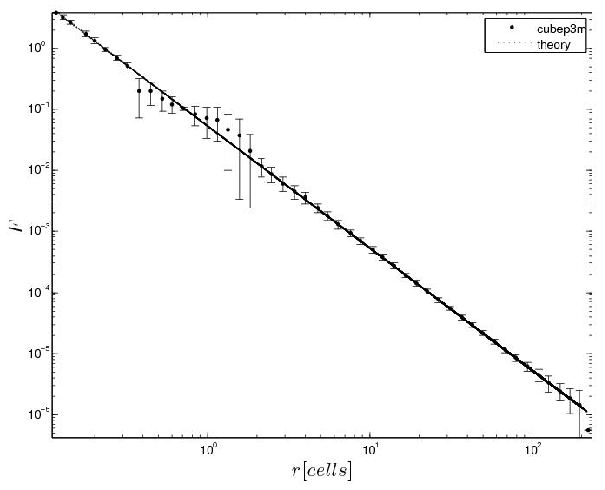}
   \includegraphics[width=3.0in]{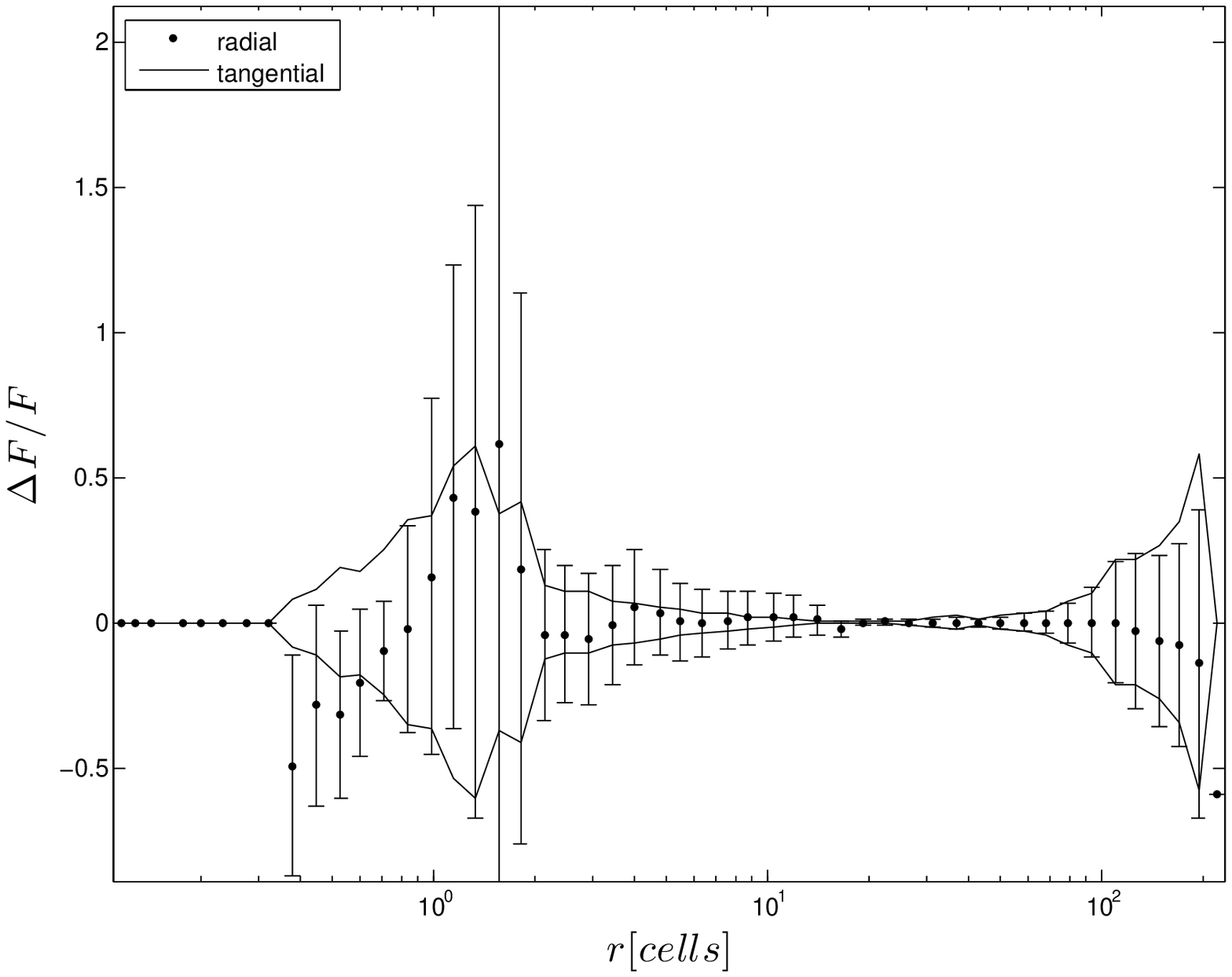}
   \includegraphics[width=3.0in]{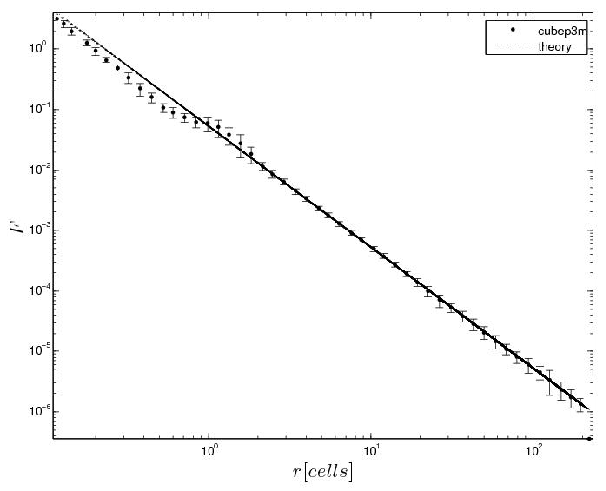}
   \includegraphics[width=3.0in]{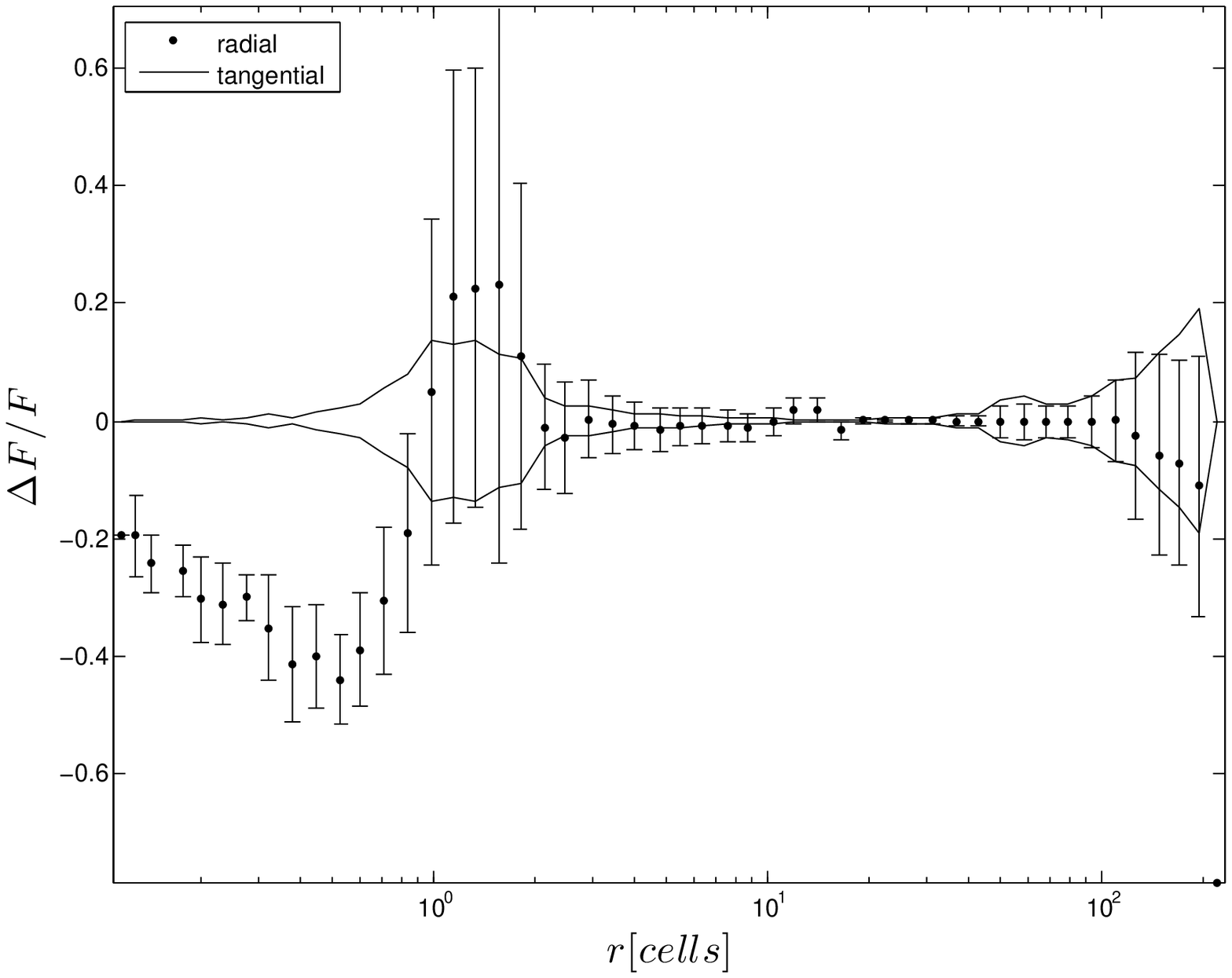}
  \caption{({\it top left}:) Gravity force in the P$^3$M algorithm, versus distance, compared with the exact $1/r^{2}$ law.
   Distance is in fine mesh cell units, and force in simulation units.
    This particular calculation was obtained in a CITA128 realization with  a box size of $100 h^{-1}\mbox{Mpc}$,
    in a single time step. 
    ({\it top right}:) Fractional error on the force in the radial direction (top) and fractional tangential contribution (bottom).
  In a full simulation run, the scatter averages over many time steps, 
    thanks to the inclusion of a random offset that is imposed on each particle, as discussed in the main text.
    ({\it middle row}:) Top row reorganized in 50 logarithmic bins; for the tangential contribution,  we plot only the scatter about the mean (solid line).  
    ({\it bottom row}:) Same as middle row, but averaged over ten time steps.
    \label{fig:den_force_fracErr}}
\end{center}
\end{figure*}

The pairwise force test presented in MPT was carried by placing two particles at random locations on the grid, calculating the force between them, then iterating
over different locations. This test is useful to quantify the accuracy on a cell-by-cell basis, but lacks the statistics that occur in a real time step calculation.  
The actual force of gravity in the P$^3$M algorithm,
as felt by a single particle during a single step, is presented in the top left panel of Fig. \ref{fig:den_force_fracErr}.
This force versus distance plot was obtained from a CITA128 realization, and the calculation proceeds in two steps: 
1- we compute the force on each particle in a given time step.
2- we remove a selected particle, thus creating a `hole', compute the force again on all particles, and record on file the 
force difference (before  and after the removal) as a function of the distance to the hole.

Particles in the same fine cell as the hole follow the exact $1/r^{2}$ curve. The scatter at 
  distances of the order of the fine grid is caused by the NGP interpolation scheme:
  particles in adjacent fine cells can be actually very close, as seen in the upper left region of this plot,
  but still feel the same mesh force at grid cell distances.
This  creates a discrepancy up to an order of magnitude, in loss or gain, depending on the location of the pair with respect to the centre of the cell.
As the separation approaches a tenth of the full box size or so, the force on the coarse mesh 
scatters significantly about Newton's law due to periodic boundary conditions. 
As mentioned at the end of section \ref{sec:Poisson}, the longest range of force kernel can either model accurately Newtonian gravity,
or the structure growth of the largest modes, but not both. For the sake of demonstrating the accuracy of the force calculation,
we chose the first option in this section, but this is not how the code would normally be used.

The top right panel of Fig. \ref{fig:den_force_fracErr} shows the fractional error on the force along the radial direction (top) and the fractional 
tangential contribution (bottom), also calculated from a single time step.
Again, particles that are in the same cell have zero fractional error in the radial direction, and zero tangential force.
Particles feel a large scatter from the neighbouring fine mesh cells, but as the distance increase beyond five fine cells, the fractional error drops down to about 20 per cent.
%The fractional error of the radial component is up to 60 per cent at sub-grid distances, but remains under ten per cent for  distances larger than a few cells. 
The transverse fractional error peaks exactly at the grid distance,  and is everywhere about 50 per cent smaller than the radial counterpart.

Although these scatter plots show the contributions from individual particles and cells, it is not clear
whether the mean radial force felt is higher or lower than the predictions. For this purpose, we rebin the results
in 50 logarithmically spaced bins and compute the mean and standard deviation. 
We show the resulting measurement in the middle left panel of Fig. \ref{fig:den_force_fracErr}, where we observe that on average,
there is a 50 per cent fractional error in the range $ 0.4 < r < 2$, but the numerical calculations are otherwise
in excellent agreement with Newton's law.
Since  the transverse force is, by definition, a positive number,
and since we know that it averages out to zero over many time steps, we plot only the scatter about the mean, represented in the figure by the solid line.
The transverse scatter is smaller everywhere for $r<10$.

At grid scales, the fractional error is larger than {\small PMFAST}, largely due to the fact that the fine mesh force is performed with an NGP interpolation scheme -- as opposed to CIC. This prescription is responsible for the larger scatter about the theoretical value, but, as mentioned earlier, 
NGP interpolation is essential to our implementation of the pp part.
At the same time, the biggest problem with the straightforward pp force calculation is that the results 
are anisotropic and depend on the location of the fine mesh with respect 
to the particles. As an example, consider two particles on either side of a grid 
cell boundary, experiencing their mutual gravity attraction via the fine mesh force with a discretized one-grid cell separation.
 If, instead, the mesh was shifted such that they were
within the same cell, they would experience the much larger pp force. 
This is clearly seen in the top left panel of Fig. \ref{fig:den_force_fracErr}, where particles physically close, but in different grid cells, 
feel a force up to a factor of five too small.
This effect is especially pronounced at the early stages of the simulation where
the density is more homogeneous, and leads to mesh artefacts appearing
in the density field.

In order to minimize these two systematic effects -- the large scatter and the anisotropy -- 
we randomly shift the particle distribution relative to the mesh by a small
amount -- up to 2 fine grid cells in magnitude -- in each
dimension and at each time step.  This adds negligible computational
overhead as it is applied during the particle position update,
and suppresses the mesh behaviour that otherwise grows over multiple time steps.
It is possible to shift back the particles at the end of each time step,
which prevents a random drift of the whole population, a necessary step 
if one needs to correlate the initial and final positions of the particles for instance,
or for hybrid dark matter -- MHD simulations.
 
We ensure that, on average, this solution balances out the mesh feature,
by tuning the force kernels such as to provide a force as evenly balanced as possible, both at the cutoff length ($r_{c}=16$ fine cells) 
and at grid cell distances (see discussion in section \ref{sec:Poisson}).
Both of these adjustments are performed from the `hole' and the pairwise force tests mentioned above.
The bottom panels of Fig. \ref{fig:den_force_fracErr} show the effect of averaging the force calculation over ten time steps;
the random offset is applied on the particles (and on the hole) at the end of each force calculation. 
We observe that the agreement with Newton's law is now at the per cent level for  $r \ge 1.5$. 
At smaller distances, we still observe a scatter, but it is about two times smaller than during a single time step for $r > 1 $.
The force at sub-grid distances is, on average, biased on the low side by 20-40 per cent,
as caused by the discretization effect of the NGP.
%{\bf (Can't we beat this by tempering with the fine force kernel? Will the least square fit enhance this?)}

We note that this inevitable loss of force in the NGP scheme is one of the driving arguments to extend the pp calculation outside the fine mesh cell,
since the scattering about the actual $1/r^{2}$ law drops rapidly with the distance.
As discussed in section \ref{subsec:extendedpp}, this gain in accuracy comes at a computational price,
but at least we have the option to run the code in a higher precision mode.

\begin{figure}%[ht]
  \begin{center}
    \includegraphics[width=3.2in]{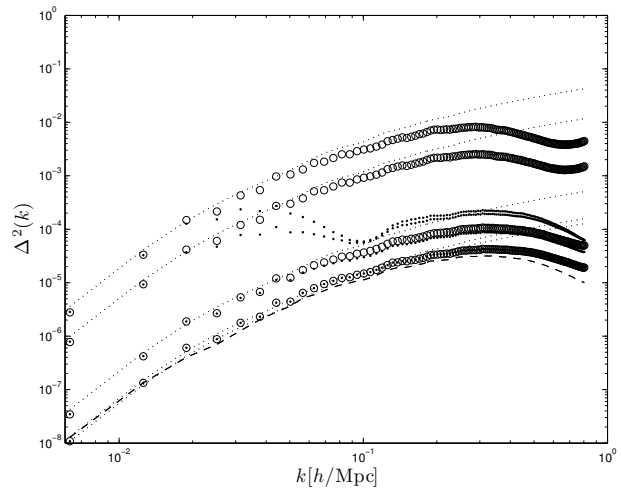}
  \caption{Dark matter power spectrum, measured at $z=180$, $100$, $20$ and $10$, in a CITA256 realization that is $1000 h^{-1}$ Mpc per side. 
  The dashed line represents the initial condition power spectrum, the dotted lines are the linear predictions, and  the open circles the standard P$^3$M configuration. 
  The dots are obtained by simply removing the random offset that is normally applied at each time step. \label{fig:disp_mesh}}
\end{center}
\end{figure}

We present in Fig. \ref{fig:disp_mesh} the dramatic impact of removing the random offset in the otherwise default code configuration.
This test was performed with a CITA256 simulation of very large box size,
the output redshifts are very early (the upper most curve at $z=10$ was obtained after
only 60 time steps), such that the agreement with linear theory should extend up to the resolution limit.
Instead, we observe that the power spectrum grows completely wrong, due to the large scatter in the force from the fine mesh,
and to the anisotropic nature of the P$^3$M calculations mentioned above.
When these effects are not averaged over, the errors directly add up at each time step,
which explains why later times are worst.
We recall that {\small PMFAST} did not have this problem since it used CIC interpolation on both meshes.  
%{\bf (Anything else to add here?)}

%\begin{figure*}%[ht]
 % \begin{center}
  %   \includegraphics[width=3.2in]{./densityForce_Zoom_N1.eps}
   % \includegraphics[width=3.2in]{./densityForce_Zoom_N10.eps}
  % \includegraphics[width=3.2in]{./densityForce_fracErr_N1.eps}
  %  \includegraphics[width=3.2in]{./densityForce_fracErr_N10.eps}
 % \caption{({\it top}:) Gravity force in of the P$^3$M algorithm, versus distance in fine mesh cell units, compared with the exact $1/r^{2}$ law.
 %   This particular calculation was obtained in a CITA128 realization with  a box size of $100 h^{-1}\mbox{Mpc}$.
  %  ({\it bottom}:) Fractional error on the force in of the P$^3$M algorithm, in the radial (top) and tangential (bottom) directions.
   % The left panels are calculated from a single time step, while the right panels show the average over time time steps.
 %   \label{fig:den_force_N10}}
%\end{center}
%\end{figure*}

%\begin{figure}%[ht]
 % \begin{center}
  %   \includegraphics[width=3.2in]{./densityForce_ppext=10_rebin.eps}
  %\caption{ Gravity force in of the P$^3$M algorithm, versus distance in fine mesh cell units, compared with the exact $1/r^{2}$ law.
   % This particular calculation was obtained in a CITA128 realization with  a box size of $100 h^{-1}\mbox{Mpc}$, averaged over 10 time steps.
   % Results are rebined to show how well the mean value follows the thoretical curve down to the grid size.
   % \label{fig:F_rebin}}
%\end{center}
%\end{figure}

\begin{figure}%[ht]
  \begin{center}
    \includegraphics[width=3.0in]{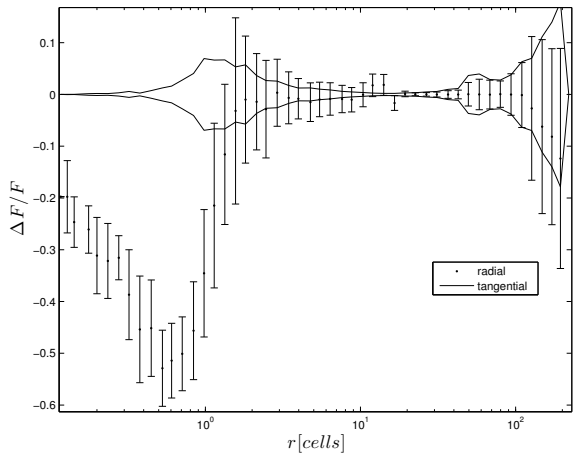}
  \caption{Fractional error in the gravity force, without the fine force kernel correction at $r=1$.
      This particular calculation was obtained in a CITA128 realization with  a box size of $150 h^{-1}\mbox{Mpc}$,
     averaged over ten time steps.
    \label{fig:den_force_fracErr_kern}}
\end{center}
\end{figure}

As mentioned in section \ref{sec:Poisson}, most of the tests presented in this paper were obtained without the fine force kernel correction near  $r=1$.
For completeness, we present here the motivation behind the recent improvement.% that compensates for a systematic underestimate at grid distances. 
Without the correction, the force of gravity looks like Fig. \ref{fig:den_force_fracErr_kern}, 
where we observe that the fractional error in the radial direction is systematically on the low side, by up to 50 per cent, even after averaging over 10 time steps. 
This is caused by the fact that in the strong regime of a $1/r^2$ law, the mean force from a cell of uniform density, as felt by a particle, 
is not equivalent to that of the same cell collapsed to its centre -- the actual position of equivalence would be at a closer distance.
Basically, the idea is to boost the small distance elements of the NGP kernel by a small amount,
which thereby corrects for the loss of force observed in Fig. \ref{fig:den_force_fracErr_kern}.

\begin{figure}%[ht]
  \begin{center}
    \includegraphics[width=3.2in]{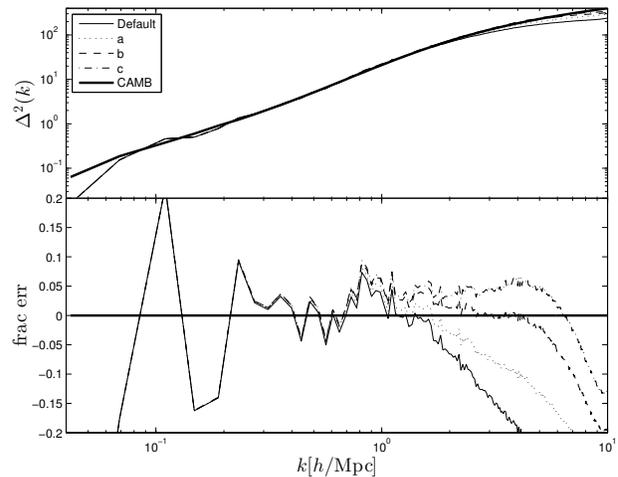}
  \caption{({\it top}:) Dark matter power spectrum, measured at $z=0$, in a SCINET256 realization that is $150 h^{-1}$ Mpc per side. 
  The thick solid line represents predictions from {\small HALOFIT}, the thin solid line is the default (i.e. without correction) configuration of the code, and the three other lines represent
  three different corrections that are applied at grid distances to compensate for the systematic force underestimation seen in the bottom panels of Fig. \ref{fig:den_force_fracErr}. 
  ({\it bottom}:) Fractional error with respect to the predictions.
  \label{fig:fudge}}
\end{center}
\end{figure}

We  show in Fig. \ref{fig:fudge} the impact on the power spectrum of three different correction trials: a) the force from the first layer of neighbours is boosted by $90$ per cent, b)
the force from the first two layers of neighbours is boosted by $60$ per cent, and c) the force from two layers is boosted by $80$ per cent.
This test was performed in a series of SCINET256 simulations with a box length of $150 h^{-1} \mbox{Mpc}$ evolved until $z=0$.
We see from the figure that in the original configuration, the departure from the non-linear predictions occur at $k\sim 1.5 h\mbox{Mpc}^{-1}$;
with the trial `a', the turn around occurs roughly at the same scale but is less steep; trial `b'  allows for a resolution gain up to $k\sim 4.5 h\mbox{Mpc}^{-1}$
before hitting a rapid drop in resolution; the last correction scheme is more aggressive, and although it recovers power at even smaller scales, 
trial `c'  exhibits a 5 per cent overestimate compared to non-linear predictions in the range $1.0 < k < 5.0 h\mbox{Mpc}^{-1}$, 
and more testing should be done prior to using this correction. 
We have also tried to increase even more the boost factor in the case a) and b), but once again, the boost started to propagate to larger scales, 
an effect we need to avoid. %We show in Fig. \ref{fig:den_force_fracErr_kern} the impact of the trial a) on the force of gravity. 
%We observe that when averaged over ten time steps, the scatter  is five to ten times larger than with the original kernel, 
%both in the radial and tangential directions, but the radial part is better balanced about the theoretical prediction. 
We therefore conclude that the fine force kernel should be subject to a boost factor following `trial b', i.e. 
$\mbox{\boldmath $\omega$}_{s}({r\le2}) \rightarrow 1.6\mbox{\boldmath $\omega$}_{s}({r\le2})$.

It is thinkable that other corrections could outperform this one, hence the kernel correction factor
should also be considered as an adjustable parameter:
`no correction' is considered as the conservative mode, which is accurate only up to the turn around 
observed in the dimensionless power spectrum, and which has been used in most of the tests presented in this paper. 
Since higher resolution can be achieved at no cost with the corrected fine force kernel, it has now been adopted in the default configuration;
trial `c' should be considered as too aggressive. We note, again, that we have not exhausted the list of possible correction,
and that future work in that direction might result in even better resolution gain.

%We note that these trials have not yet been 
%tried on very large runs, and we recommend that some more testing should be done to assess the robustness of this correction. 

\subsection{Constraining redshift jumps}
\label{subsec:z_jumps}

At early stages of the simulation, the density field is homogenous, causing the force of gravity to be
rather weak everywhere. In that case, the size of the redshift jumps is controlled by a limit in the cosmological expansion.
If the expansion jump is too large, the size of the residual errors can become significant, and one can observe, for instance,
a growth of structure that does not match the predictions of  linear theory even at the largest scales.
One therefore needs to choose a maximum step size. In {\small CUBEP$^3$M}, this is controlled by $ra_{max}$, which is the maximal fractional step size,
$\mbox{d}a/(a + \mbox{d}a)$, and is set to $0.01$ by default.  Generally, a simulation should start at a redshift high enough so that
the initial dimensionless power spectrum is well under unity at all scales, ensuring that the Zel'dovich approximation
 holds at the per cent level at least. At the same time, starting too early yields particle displacements and velocities that are too small;
 in that case, truncation error at the early time steps can be significant, causing a drop of accuracy.

\begin{figure}%[ht]
  \begin{center}
    \includegraphics[width=3.2in]{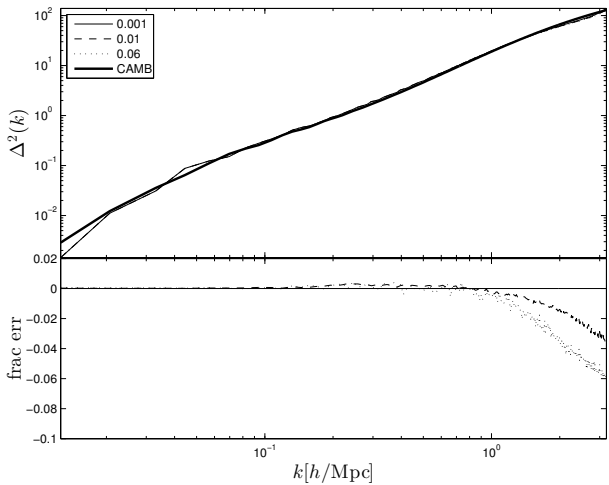}
  \caption{({\it top}:) Dark matter power spectrum, measured at $z=0$ in a series of SCINET256 realizations, started at $z_i = 100$. 
 The different curves show different values of $ra_{max}$. 
  The resources required to run these simulations increase rapidly as $ra_{max}$ decreases, as seen in Table \ref{table:ra_max}.
  ({\it bottom}:) Fractional error with respect to the most accurate run, i.e. with $r_{max} = 0.001$. \label{fig:ra_max}}
\end{center}
\end{figure}

It is possible to reduce this effect, and thereby improve significantly 
the accuracy of the code, by decreasing the value of $ra_{max}$, at the cost of increasing the total number of time steps.
Fig. \ref{fig:ra_max} shows a comparison of late time power spectra of a series of SCINET256 realizations that originate from the same initial conditions, 
and used the same random seeds to control the fine mesh shifts (mentioned above): only the value of $ra_{max}$ was modified between each run. 
We observe that the impact is mostly located in the non-linear regime, where decreasing the time step from $0.06$
to $0.001$ allows the simulation to recover about $6$ per cent of dimensionless power at the smallest scale.
In this test case, reducing at the intermediate value of $ra_{max} = 0.01$ shows a $\sim 4$ per cent accuracy loss over the $0.001$ case, 
and the increase in time is 14 per cent  compared to the $0.06$ case.
We have also run a $ra_{max} = 0.0005$ simulation, which showed sub percent changes even at the largest $k$.
This confirms that the results at $0.001$ have converged for the setup under study 

Again, this gain/loss of power is greatly dependent on the choice of initial redshift, the resolution and the box size, and ideally one would make
test runs in order to optimize a given configuration.  
As expected, the {\small CPU} resources required to run these simulations increase quite rapidly as $ra_{max}$ decreases, as seen in Table \ref{table:ra_max}. 
The default configuration of the code was previously set to $ra_{max}=0.05$, but in the light of this recent test, 
the it has been set to $0.01$,  with the option to run at $0.001$ ($0.05$) for higher (lower) accuracy mode.

%We emphasize again that this is only a safety net for simulations that start at very high redshifts, and we have checked that it has negligible impact 
%on simulations that started at later redshifts. 

\begin{table}
\begin{center}
\caption{Scaling in {\small CPU} resources as a function of the value of $ra_{max}$. The tests were performed 
on the SciNet GPC cluster, and general trends could vary slightly on other machines. The time tabulated on the right column
corresponds to the full run time of the SCINET256 simulations, which evolved $256^{3}$ particles from $z = 100$ down to $z = 0$.}
\begin{tabular}{|l|c|c|}
\hline 
$ra_{max}$         & time (h)   \\                 
\hline
 $0.06$ & 3.11\\
 $0.01$ & 3.56 \\
 $0.006$ & 4.16\\
 $0.003$ & 6.01 \\
 $0.002$ & 8.65\\
 $0.001$ & 17.49\\
\hline
\end{tabular}
\label{table:ra_max}
\end{center}
\end{table}

%{\bf (Any other systematics effects we want to discuss here?)}

%\input{../cubep3m_paper/Halos}

\section{Runtime Halo Finder}
\label{sec:halo}

%Spherical overdensities, search algorithm, 
%provided halo information, halo bias, comparison to PS and ST.
%{\bf (Ilian, You can lead the way here...)}

We have implemented a runtime halo finding procedure, which we have developed 
based on the spherical overdensity (SO) approach \citep{1994MNRAS.271..676L},
which we refer to as the `CPMSO' halo finder (for CubeP$^{3}$M SO).
In the interest of speed and efficiency, the halo catalogues are constructed 
on-the-fly at a pre-determined list of redshifts. The halo finding is 
massively-parallel and threaded based on the main {\small CUBEP$^3$M} data structures 
discussed in section \ref{sec:structure}. 

From the particle list, the code first builds the 
fine-mesh density for each sub-domain, using either CIC or NGP interpolation. It then 
proceeds to search for all local maxima above a certain
threshold (typically set to a factor of 100 above mean density) within each local density tile. 
It then uses parabolic interpolation on the density field to determine more precisely
the location of the maximum within the densest cell, and records the peak 
position and value. %The halo centre determined this way agrees closely with 
%the centre-of-mass of the halo particles discussed below.  

Once the list of peak positions is generated, they are sorted from the highest 
to the lowest density. Each halo candidates is then inspected 
independently, starting with the highest peak. The grid mass is accumulated 
in spherical shells of fine grid cells surrounding the maximum, until the 
mean density within the halo drops below a pre-defined overdensity cutoff 
(usually set to 178 in units of the mean, in accordance with the top-hat 
collapse model). As we accumulate the mass, we remove it from the mesh, so that no 
element is double-counted. This method is thus inappropriate for finding 
sub-haloes since, within this framework, those are naturally incorporated in their 
host haloes. Because the haloes are found on a discrete grid, it is 
possible, especially for those with lower mass, to overshoot the target overdensity.
To minimize this effect, we correct the halo mass and radius with an analytical density profile. 
We use the Truncated Isothermal Sphere (TIS) 
profile \citep{1999MNRAS.307..203S,2001MNRAS.325..468I} for overdensities below 
$\sim130$, and a simple $1/r^2$ law for lower overdensities. 
The TIS profile yields a similar outer slope to the Navarro, Frenk and White 
\citep[NFW][]{1997ApJ...490..493N} profile, but extends to lower overdensities
and matches well the virialization shock position given by the
self-similar collapse solution of \citet{1985ApJS...58...39B}.

After the correct halo mass, radius and position are determined, we find all 
particles that are within the halo radius. Their positions and velocities are
used to calculate the halo centre-of-mass, bulk velocity, internal velocity 
dispersion and the three angular momentum components, all of which are then 
included in the final halo catalogues. We also calculate the total mass of
all particles within the halo radius, also listed in the halo data. This mass
is very close, but  typically slightly lower by a few per cent, than the halo mass calculated 
based on the gridded density field. The  centre-of-mass found this way closely follows 
that found from the peak location, which is based on the gridded mass distribution. 
However, it is common for halos that have underwent recent merging to have 
their peak location further away from their centre of mass.

%Compared to Tinker
%44% low for 50 particles
%22% low for 400 particles
%12% low for 1000 particles

\begin{figure}%[ht]
%  \vskip -0.5cm 
  \begin{center}
    \includegraphics[width=3.2in]{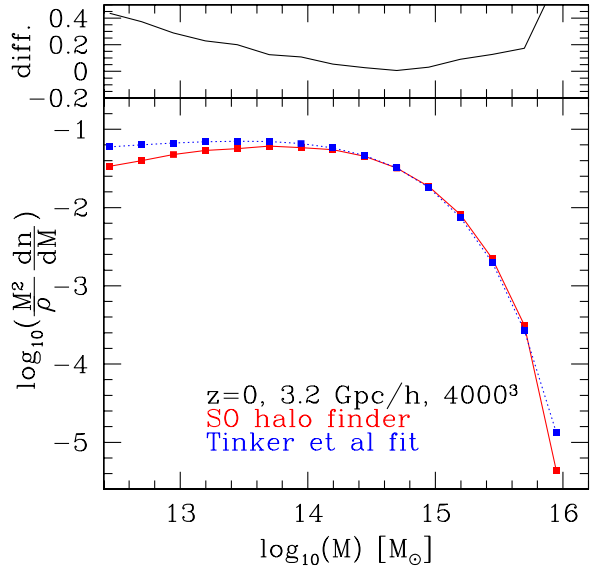}
  \caption{({\it main} : ) Simulated halo multiplicity function, 
    $\frac{M^2}{\bar{\rho}}\frac{dn}{dM}$, based on a
    RANGER4000 simulation with $3.2\,h^{-1} \mbox{Gpc}$ box and $4000^3$ 
    particles (solid, red in the online version). For reference we also show a widely-used 
    fit by \citet{2008ApJ...688..709T} (blue, dotted). 
    The particle masses are of $5.68 \times 10^{10} M_{\odot}$.
    ({\it top} : ) Absolute value of the fractional difference with respect to the fit, i.e. $|$Fit - SO$|$/Fit.
    \label{mf}}
     \end{center}
\end{figure}

A sample halo mass function produced  from a RANGER4000 simulation at $z=0$ is shown in Fig. \ref{mf}. We compare our result to the 
precise fit presented recently by \citet{2008ApJ...688..709T}. Unlike most
other widely-used fits like the one by \citet{2002MNRAS.329...61S}, which are based on friends-of-friends (FOF)
halo finders, this one relies on the
SO search algorithm, whose masses are systematically different 
from the FOF masses \citep[e.g.][]{2007MNRAS.374....2R,2008ApJ...688..709T}, 
making this fit a better base for comparison here. Results show excellent
agreement, within $\sim10$ per cent for all haloes with masses corresponding to
1000 particles or more. The factor of two discrepancy at high mass comes from inaccuracies in the model, 
which did not include those large scales in the original fit. We refer the reader to \citet{2012arXiv1212.0095W}
for extensive details about comparison between CPMSO,  FOF and the Amiga halo finder \citep{2009ApJS..182..608K}.

Lower-mass haloes are under-counted compared
to the \citet{2008ApJ...688..709T} fit, by $\sim20$ per cent for 400 particles and 
by $\sim40$ per cent for 50 particles (not shown in the figure). 
This is largely due to the grid-based nature of our SO halo finder, which misses some of the low-mass haloes. 
More sophisticated halo finders (available only through post-processing calculations due to their heavier memory footprint) can recover a number count -- i.e. a halo mass function -- much closer to the theoretical predictions in  the regime of a  few tens of particles, as these take advantage of techniques such as mesh refinement, full 6-dimensional phase space information or even merging history  \citep[see][for a comprehensive comparison between halo finders]{2011MNRAS.415.2293K}.
%It was shown that using more sophisticated halo finders (available only through post-processing calculations due to their heavier memory
%footprint) it is possible to recover a number count much closer to the theoretical predictions down to {\bf XXXX} particles.

We could in principle run the halo finder on a finer mesh, taking advantage of the sub-grid resolution to identify more accurately the collapsed regions, 
which would certainly allow us to  identify better the small mass halos.
However, we recall here that since {\small CUBEP$^3$M} is designed to be light and fast, we paid attention that the halo finding procedure
does run with negligible extra memory. 
This choice comes at a price: in order to achieve an accuracy of a few  per cent on the halo mass function, 
the CPMSO halo finder requires about 500 particles per halo; smaller objects are under counted.
However, many halo properties require that kind of particle count to reach the required level of  accuracy, 
plus it is always possible to run a different halo finder on the {\small CUBEP$^3$M} particle dump
if one needs the smaller haloes.
%{\bf The referee thinks that we should shout this out loud, as many other on-the-fly halo finder go down to 50 particles, still with good counting... 400-1000 particles is rather poor in comparison.}

%A second test of the accuracy of the halo finder algorithm is to extract the halo power spectrum $P_{h}(k)$ and compare the halo bias
%with theoretical predictions. The halo bias is defined as $b(k) = \sqrt{P(k)/P_{h}(k)}$ and is shown in Fig. \ref{fig:halo}.
%The results are organized in four mass bins, where the lowest mass consists of 20-200 particles, the second bin 
%contains 200-2000 particles and so on. Comparisons with the linear predictions of \citet{2001MNRAS.323....1S} shows
%results consistent the known flaws of the model, namely that low-mass bias is over-predicted and high-mass bias is under-predicted \citep{2010ApJ...724..878T}.
%
%\begin{figure}%[ht]
%  \vskip -0.5cm 
%  \begin{center}
%    \includegraphics[width=3.2in]{./bias.eps}
%  \end{center}
%  \caption{Halo bias, measured at $z=0$ from a RANGER4000 simulation with a side length of $3.2 h^{-1}$Gpc.
%  Bottom to top curves correspond to haloes of different masses, from light to massive. The four bins are decades
%  of halo masses; the first contains haloes made of 20-200 particles, the second of 200-2000 and so on.
%  The dashed lines are linear predictions from \protect \citet{2001MNRAS.323....1S}.
%    \label{fig:halo}}
%\end{figure}

%{\bf Show some examples and comparisions to analytical fits and other halo finders.}

%\input{../cubep3m_paper/Extensions}

\section{Beyond the standard configuration}
\label{sec:extensions}

Whereas all the preceding sections contain descriptions and discussions that apply to the standard configuration of the code, 
many extensions have been recently developed in order to enlarge the range of applications of {\small CUBEP$^3$M};
this section briefly describes the most important of these improvements.

\subsection{Initial conditions}
\label{subsec:init}

As mention in section \ref{sec:structure}, the code starts off by reading a set of initial conditions.
These are organized as a set of $6 \times N$ phase-space arrays -- one per {\small MPI} process -- where $N$ is the number of particles in the
local volume. Each {\small MPI} sub-volume generates its own random seed, 
which is used to create a local noise map. The initial power spectrum is found by reading off a transfer function at redshift $z=0$ that is generated from {\small CAMB}  by default, 
and evolved to the desired initial redshift with the linear $\Lambda$CDM growth function. 
The noise map is then Fourier transformed, each element is multiplied by the (square root of the) power spectrum, and the result is brought back to real space. 
To ensure that the global volume is correct, a buffer layer in the gravitational potential is exchanged across each adjacent node.
The particles are then moved from a uniform cell-centred position to a displaced position with Zel'dovich approximation\footnote{We remind that 
for Zel'dovich approximation to hold in such cases, the simulations need to be started at very early redshifts.
Consequently, the size of the first few redshift jumps in such simulations can become rather large, and therefore less accurate, 
which is why we must carefully constrain the size of the jumps, as discussed in section \ref{subsec:z_jumps}.
Another solution to this issue is to use second order perturbation theory to generate the initial conditions,
which is not implemented yet in the public package.}.

The density field produced by this algorithm follows Gaussian statistics, and is well suited to describe many systems.
To increase the range of applicability of this code, we have  extended this Gaussian initial conditions generator
to include non-Gaussian features of the `local' form,
$\Phi({\bf x})=\phi({\bf x})+f_{\rm NL} \phi({\bf x})^2 + g_{\rm NL} 
\phi({\bf x})^3$, where $\phi({\bf x})$ is the Gaussian contribution
to the Bardeen potential $\Phi({\bf x})$ (see \cite{2004PhR...402..103B} for a review). 
We adopted the CMB convention,
in which $\Phi$ is calculated immediately after the matter-
radiation equality (and not at redshift $z=0$ as in the large scale
structure convention). For consistency, $\phi({\bf x})$ is normalized
to the amplitude of scalar perturbations inferred by CMB measurements
($A_s\approx 2.2 \times 10^{-9}$). The local transformation is performed 
before the inclusion of the matter transfer function, and the initial 
particle positions and velocities are finally computed from $\Phi({\bf x})$ 
according to the Zel'dovich approximation, as in the original Gaussian initial condition generator.

\begin{figure}%[ht]
  \begin{center}
    \includegraphics[width=3.2in]{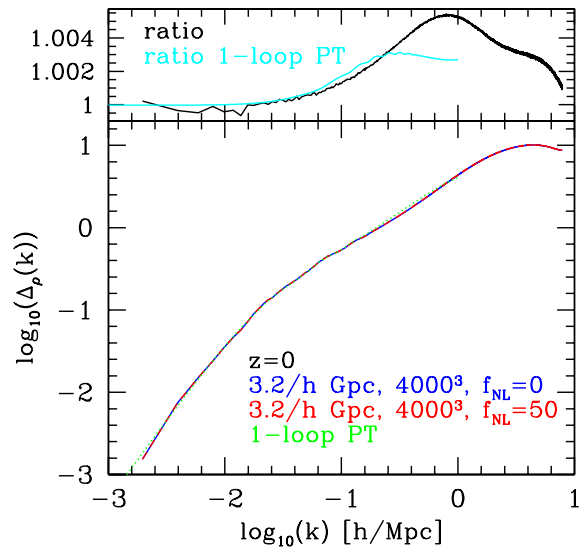}
  \caption{ Dark matter power spectrum, measured at $z=0$ in a volume $3.2 h^{-1}\mbox{Gpc}$ per side,
  from $4000^3$ particles. The two curves represent two RANGER4000 realizations of the same initial power spectrum, one of which used Gaussian statistics (blue in the online version) and the other the non-Gaussian initial condition generator (red online). The two curves differ at the sub-per cent level, as seen in the top panel,
  and the one-loop perturbation theory calculations (aqua online) accurately describes the ratio between the two curves up to $k\sim 0.4h\mbox{Mpc}^{-1}$ in this case.   
    \label{fig:init}}
\end{center}
\end{figure}

This code was tested by comparing simulations and theoretical predictions
for the effect of local primordial non-Gaussianity on the halo mass 
function and matter power spectrum (Desjacques, Seljak \& Iliev 2009). 
It has also been used to quantify the impact of local non-Gaussian initial
conditions on the halo power spectrum \citep{2009MNRAS.396...85D,
2010PhRvD..81b3006D} and bispectrum \citep{2010MNRAS.406.1014S},
 as well as the matter bispectrum \citep{2011arXiv1111.6966S}.
Fig. \ref{fig:init} shows the late time power spectrum of two RANGER4000 realizations that started off the same initial power spectrum, 
but one of which had non-Gaussian features set to $f_{NL} = 50$.
We see that the difference between the two power spectra is at the sub-per cent level, and that the ratio of the two power spectra
is well described with one loop perturbation theory \citep{2004PhRvD..69j3513S,2008PhRvD..78l3534T}.

\subsection{Particle identification tags}
\label{PID}

A system of particle identification can be turned on, which basically allows to track each particle's trajectory
between checkpoints. Such a tool is useful for a number of applications, from reconstruction of halo merging history to tracking individual particle
trajectories.
The particle tag system has been implemented as an array of double precision integers, {\tt PID}, 
and assigns a unique integer to each particle during the initialization stage. 
Since the array takes a significant amount of memory, we opted to store the tags in a separate object
-- as opposed to adding  an extra dimension to the existing {\tt xv} array, such that it can be turned off 
when memory becomes an issue. Also, it simplifies many post-processing applications that read the position-velocity only.
The location of each tag on the {\tt PID} array matches the location of the corresponding particle on the {\tt xv} array, hence in practice it acts just like an extra dimension.
The tags on the array change only when particles exit the local volume, in which case the tag is sent along the particle to adjacent nodes 
 in the {\tt pass\_particle} subroutine; similarly, deleted particles result in deleted tags.
As for the {\tt xv} arrays, the {\tt PID} arrays get written to file in parallel at each particle checkpoint.

\subsection{Extended range of the pp force}
\label{subsec:extendedpp} 

As discussed in section \ref{subsec:force}, one of the main sources of error in the calculation of the force occurs from the PM interaction at the smallest scales of the fine grid.
The NGP approximation is the least accurate there, which causes a maximal scatter about the exact $1/r^2$ law.
A straightforward solution to minimize this error consists in extending the pp force calculation outside a single cell,
up to a range where the scatter is smaller.
Although this inevitably reintroduces a number of operations that scales as $N^2$, 
our goal is to add the flexibility to have a code that runs slower, but produces results with a higher precision. 

To allow this feature, we have to choose how far outside a cell we want the exact pp force to be active.  
Since the force kernels on both meshes are organized in terms of grids, the simplest way to implement this 
feature is to shut down the mesh kernels in a region of specified size, and allow the pp force to extend therein.
Concretely, these regions are constructed as cubic layers of fine mesh grids around a central cell; 
the freedom we have is to choose the number of such layers.
 
\begin{figure*}%[ht]
  \begin{center}
    \includegraphics[width=3.0in]{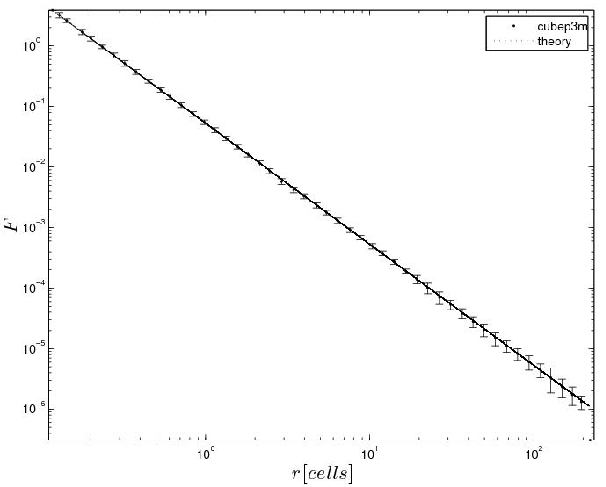}
    \includegraphics[width=3.0in]{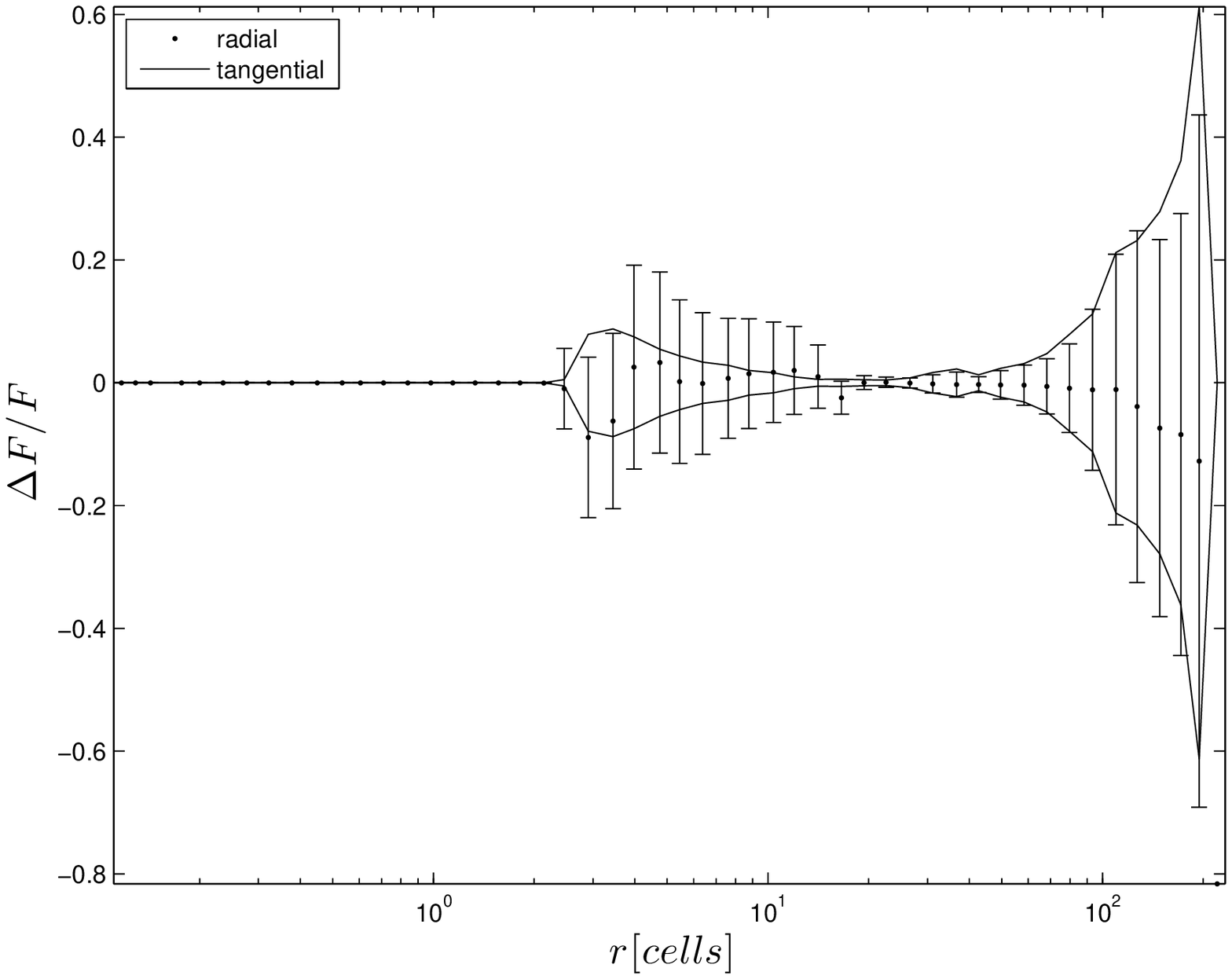}
     \includegraphics[width=3.0in]{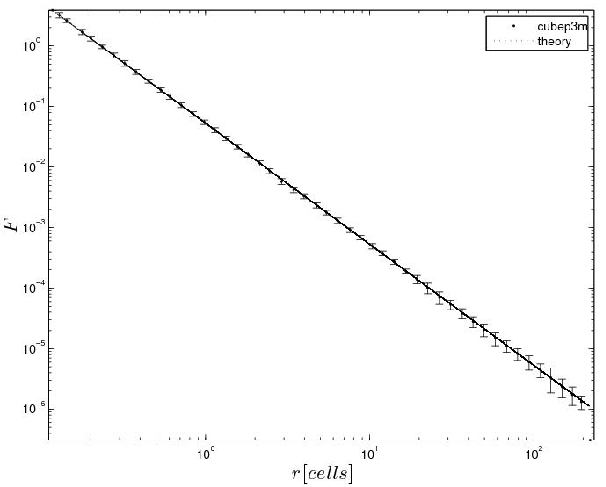}
    \includegraphics[width=3.0in]{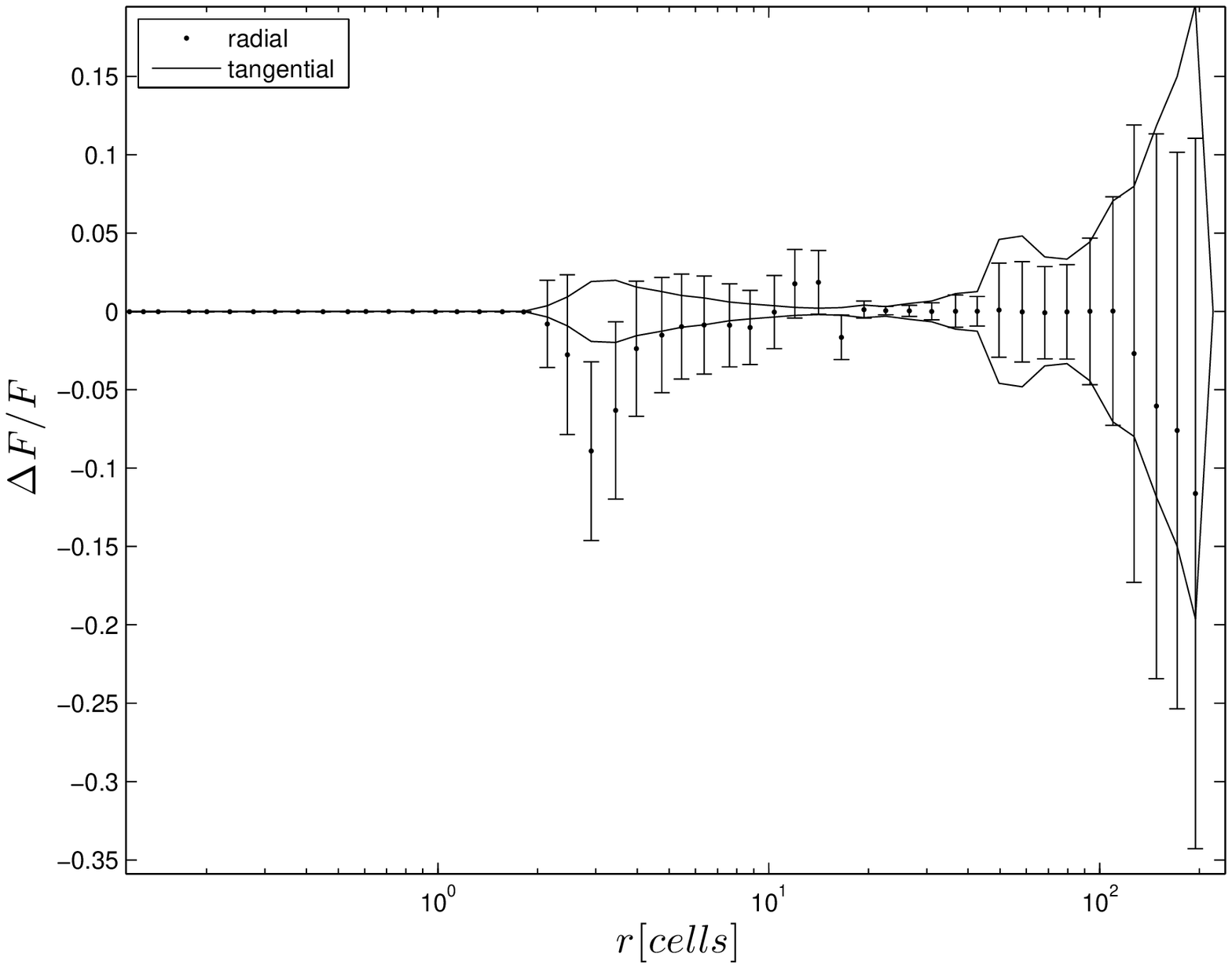}
    \caption{({\it top left}:) Gravity force in the P$^3$M algorithm, compared with the exact $1/r^{2}$ law,
  in the same CITA128 realization as that shown in Fig. \ref{fig:den_force_fracErr}, 
  except that the exact pp force has been extended to two fine mesh layers around each particle
  in the force test code.
  Particles in that range follow the exact curve, after which we observe a scatter at 
  distances of the order of 2 fine grid that is much smaller than that observed in Fig. \ref{fig:den_force_fracErr}. 
  The NGP interpolation scheme is again responsible for the scatter, but the effect is suppressed for increasing distances.
  %\caption{
  ({\it top right}:) Fractional error on the force in the radial direction (points), over plotted with the scatter in the fractional tangential contribution (solid line).
  This was also obtained over a single time step.
  ({\it bottom row}:) Same as top row, but averaged over ten time steps.
    \label{fig:den_force_fracErr_ppext6}}
\end{center}
\end{figure*}

 To speed up the access to all particles within the domain of computation, we construct a thread safe linked list
 to be constructed and accessed in parallel by each core of the system, but this time with a head-of-chain that points to the first particle in the current fine mesh cell. 
 We then loop over all fine grids, accessing the particles contained therein and inside each fine grid cell for which we killed the mesh kernels,
 we compute the separation and the force between each pair and update their velocities simultaneously with Newton's third law. 
 To avoid double counting, we loop only over the fine mesh neighbours that produce non-redundant contributions. Namely, for a central cell located at 
 $(x_1, y_1, z_1)$, we only consider the neighbours $(x_2, y_2, z_2)$ that satisfy one of the following conditions:
 \begin{itemize}
% \item{always count if $z_2 \ge z_1$ }
 %\item{if $z_2 = z_1$, then $y_2 \ge y_1$, otherwise we also allow $y_2 < y_1$} 
 %\item{if $z_2 = z_1$ and $y_2 = y_1$, then we enforce $x_2 > x_1$}
 \item{$z_2 \ge z_1$ }
 \item{$z_2 = z_1$ and $y_2 \ge y_1$} 
 \item{$z_2 = z_1$ and $y_2 = y_1$ and $x_2 > x_1$}
 \end{itemize}
 The case where all three coordinates are equal is already calculated in the standard pp calculation of the code, hence we don't recount it  here.
 To assess the improvement of the force calculation, we present in Fig \ref{fig:den_force_fracErr_ppext6} a force versus distance
 plot analogous to Fig. \ref{fig:den_force_fracErr}, but this time the pp force has been extended to  two layers of fine cells (again 
 in a CITA128 realization).
 We observe that the scatter about the theoretical curve has reduced significantly, down to the few percent level, 
 and is still well balanced around the theoretical predictions.
 The fractional error on the radial and tangential components of the force, as seen in the right panel,
 are now at least five times smaller than in the default P$^{3}$M algorithm.
 When averaging over 10 time steps, we observe that improvement is relatively mild, showing that the calculations are already very accurate. 
 
\begin{figure}
  \begin{center}
  \epsfig{file=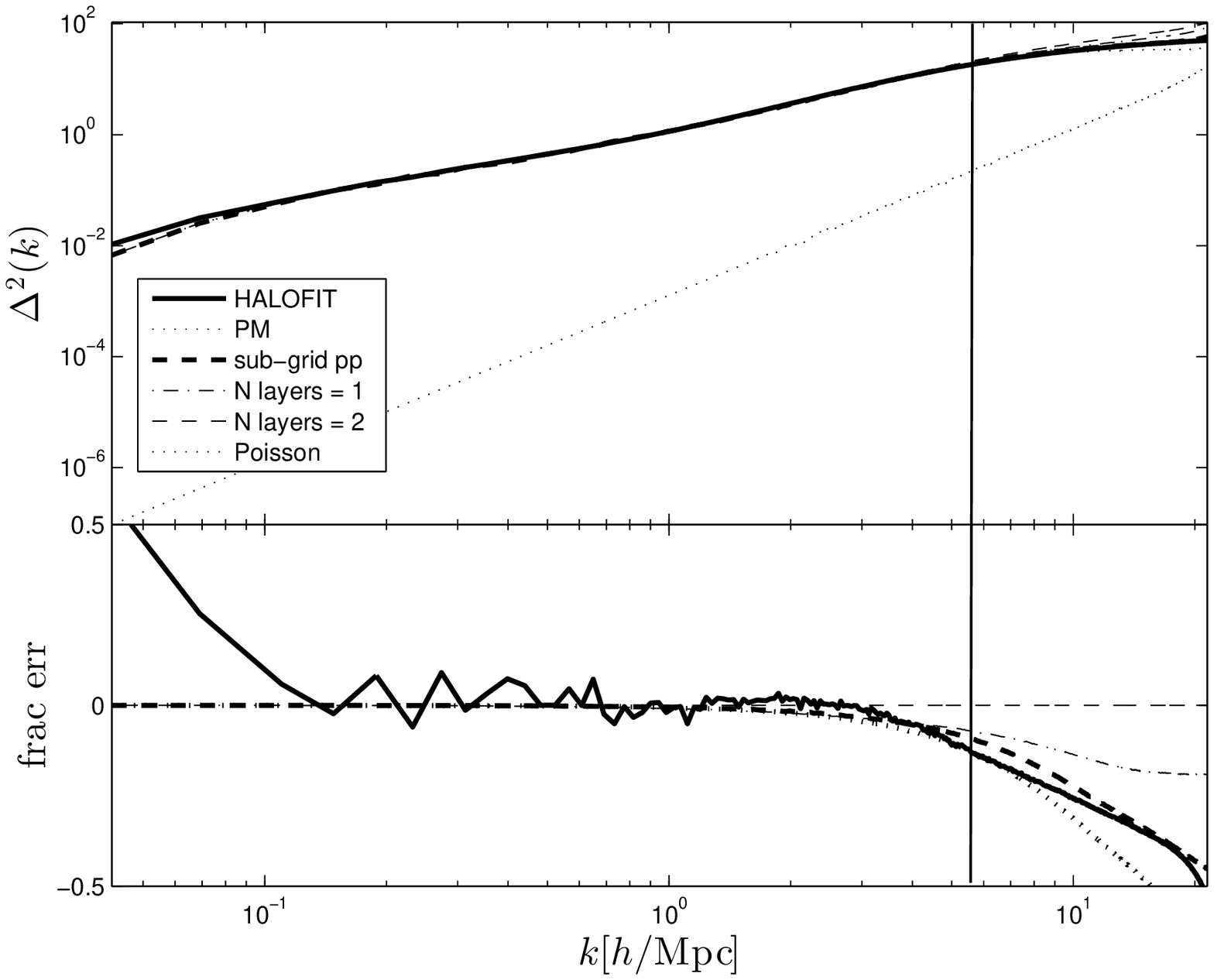,width=0.49\textwidth}
  \caption{ ({\it top}:) Dimensionless power spectrum for varying ranges of the exact pp force.
  These SCINET512 realizations evolved from a unique set of initial conditions at a starting redshift of $z = 100$ until $z=2.0$, with a box size equal to $150 h^{-1}\mbox{Mpc}$.
  The only difference between the runs is the ranges of the pp calculation. The thick vertical line corresponds to the coarse mesh scale. 
  ({\it bottom}:) Fractional error with respect to the $N_{layer} = 2$ case, which is the most accurate.}
    \label{fig:power}
  \end{center}
\end{figure}

 To quantify the accuracy improvement versus computing time requirements, we perform the following test.
 We generate a set of initial conditions at a starting redshift of $z = 100$, with a box size equal to $ 150 h^{-1}\mbox{Mpc}$,
 and with $512^{3}$ particles. We evolve these SCINET512 realizations with different ranges for the pp calculation, and compare 
 the resulting power spectra. For the results to be meaningful, we again need to use the same seeds for the random number generator,
 such that the only difference between different runs is the range of the pp force.
 Fig. \ref{fig:power} shows the dimensionless power spectrum of the different runs at $z=2.0$. We first see a significant gain in resolution
 when extending  PM to P$^{3}$M;  gains roughly as strong are found when adding successively one and two layers of fine cell mesh in which the pp force is extended.
We have not plotted the results for higher numbers of layers, as the improvement becomes milder there while the runs take increasingly more time to complete.
For this reason, it seems that a range of two layers suffices  to reduce most of the undesired NGP scatter.

Extending the pp calculation comes at a price, since the number of operations scales as  $N^{2}$ in the sub-domain. 
This cost is best captured by the increase of real time required by a fixed number of dedicated  {\small CPU}s 
to evolved the particles to the final redshift. For instance, in our SCINET512 simulations, 
the difference between the default configuration and $N_{layer} = 1$  is about a factor of $2.78$ in real time,
and about $6.5$ for $N_{layer} = 2$.
This number will change depending on the problem at hand and on the machine, and we recommend
to perform performance gain vs resource usage tests on smaller runs before running large scale simulations
with the extended pp force.

%\begin{table}
%\begin{center}
%\caption{Scaling in {\small CPU} resources as a function of the range of the pp interaction.
%$N_{layers}$ refers to the number of fine mesh layers around a given cell, inside of which the force calculation
%is purely given by the pp contribution. }
%\begin{tabular}{|l|c|c|}
%\hline 
%             Type         & time (h)   \\
%                  \hline
%PM                         & 1.77 \\
%P$^{3}$M             & 2.09 \\
%\hline
%$N_{layers}$       &          \\
%\hline
% $1$ & 8.74 \\
% $2$ & 11.47\\
% $3$ & 14.30 \\
 %$4$ & 18.87\\
 %$5$ & 22.52\\
 %$6$ & 29.62\\
 %$7$ & 34.82\\
 %$8$ & 47.00\\
%\hline \hline
%\end{tabular}
%\label{table:cpu_pp_ext}
%\end{center}
%\end{table}

\begin{figure}
  \begin{center}
  \epsfig{file=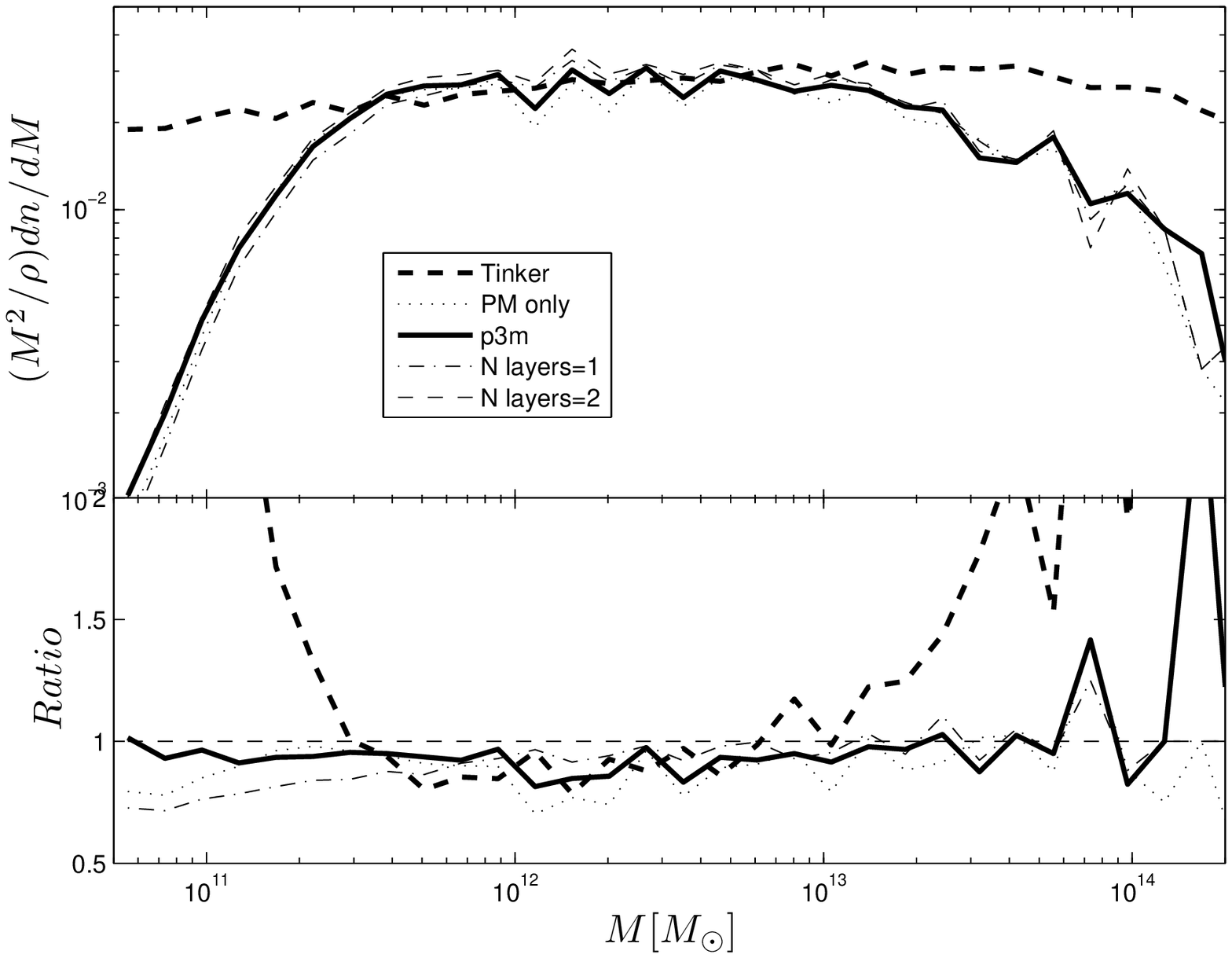,width=0.49\textwidth}
  \caption{ ({\it top}:) Halo mass function in our set of SCINET512 simulations, for different ranges of the pp force calculation, compared to the predictions of Tinker (2008), at $z = 1.0$. 
  The smallest haloes shown here have a mass equivalent to 20 particles of $2.79 \times 10^{9} M{\odot}$ each, and fall in the lowest mass bin.
  ({\it bottom}:) Ratio between the different curves and that of the $N_{layer} = 2$ configuration.}
    \label{fig:MassFunction_extpp}
  \end{center}
\end{figure}

The power spectrum does not provide the complete story, and one of the most relevant ways to quantify the improvement of the calculations is to compare the halo mass function from these different SCINET512 runs. Fig. \ref{fig:MassFunction_extpp} presents this comparison at redshift $z = 1.0$. About 76,000 haloes were found in the end, yielding a halo number density of about $0.023$ per $[\mbox{Mpc/$h$}]^{3}$. 
We observe that these simulations undershoot the Tinker (2008) predictions, compared to the RANGER4000 run, an effect caused by the much lower resolution of the configuration. 
We also observe that the pure PM code yields up to 10 per cent less haloes than the P$^{3}$M version over most of the mass range,
whereas the extended pp algorithm generally recovers  up to 10 per cent more haloes in the range $10^{11} - 10^{13} M_{\odot}$.

\begin{figure}
  \begin{center}
  \epsfig{file=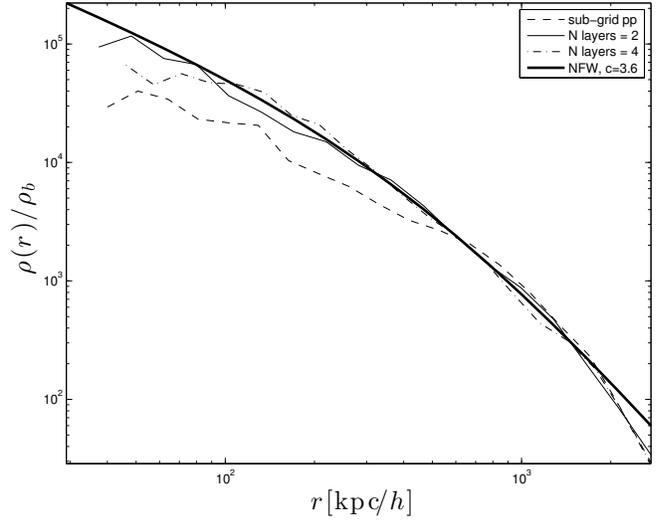,width=0.49\textwidth}
  \caption{ Halo density profile for different ranges of the pp force calculation in a CURIE432 simulation, compared to the NFW predictions, at $z = 0.026$. 
  The halo shown here has a mass of $1.1 \times 10^{15} M{\odot}$, a virial radius of   $r_{200} = 1.74 \times 10^{3} h^{-1}\mbox{kpc} $, and we used $c=3.6$ in the NFW fit.
  The difference between $N_{layer} = 2$ and $4$ is quite small, but they both outperform the sub-grid calculations.}
    \label{fig:profile}
  \end{center}
\end{figure}

A third way to assess the improvement from the extended pp force is to study its effect on halo density profiles.
Since the on-the-fly halo finder is neither equipped nor optimal to perform such calculations, we ran the {\small AMIGA} halo finder \citep{2009ApJS..182..608K} 
on a series of CURIE432 simulations, which were run on 27 nodes at the CURIE supercomputing centre.
These runs started at redshift $z_i = 100$, and evolved $432^3$ particles until $z= 0.026$ in a comoving volume of $432 h^{-1}\mbox{Mpc}$ per side. 
Fig. \ref{fig:profile} presents the radial profiles of a $1.1\times 10^{15} M_{\odot}$ halo, 
compared to the  \citet{1997ApJ...490..493N} predictions with a concentration parameter of $c = 3.6$. 
We see that the predictions are well reproduced down to $40 h^{-1} \mbox{kpc}$ when the pp force is extended to two layers of fine cells.
Here again, we observe that the difference in performance between the extension of the pp  range to two vs four cells deep is rather mild, from where
we conclude that $N_{layers} = 2$ is the choice that optimizes speed and accuracy.

\subsection{Summary of the accuracy parameters}

As we have seen, the accuracy of CUBEP3M can be modify by turning 4 knobs, 
each having their own impact, and we briefly summarize here their impact on the calculation. These four parameters are:

\begin{itemize}
\item{$r_{soft}$ (see section \ref{subsec:force_soft}),}
\item{Correction to the fine mesh force kernel (see section \ref{subsec:force}),}
\item{$ra_{max}$ (see section \ref{subsec:z_jumps}),}
\item{range of the the pp force (see section \ref{subsec:extendedpp})}
\end{itemize}

$r_{soft}$ and the range of the pp force affect the code accuracy regardless of its scale and volume, as they control the resolution, but more accurate results come the cost of larger computing time. 
The kernel correction factor, which comes at no cost at all, 
affects the force accuracy regardless of the size and volume, as it corrects for a bias 
that occur in the calculation at the  distance of $\sim 1-2$ fine cells.
The last accuracy feature, $ra_{max}$, affects the maximum time step size, 
and has an effect only at early time, when forces are small.
%In configurations where the resolution is high enough, and the initial time step is not too large, then the mesh force will already be large enough to dominate the time step budget

\section{Conclusion}

This paper describes {\small CUBEP$^3$M}, a public and  massively parallel P$^3$M N-body code that inherits from {\small PMFAST} 
 and that now scales well to 27,000 cores, pushing the limits of the cosmological problem size one can handle.
 This code is fast and has a memory footprint that can be made comparable to that of Lean-{\small GADGET-3} at peak consumption. 
The user gets to chose between extra-lean modes that can use as low as $37$ bytes per particles (plus load imbalances),
and memory heavier configurations that run much faster. 
We summarize the code structure, review the double-mesh Poisson solver algorithm, and present scaling and systematic tests
that have been performed. We also describe various utilities and extensions that come with the public release, 
including a run time halo finder, an extended pp force calculation, a system of particle ID and a non-Gaussian initial condition generator.
{\small CUBEP$^3$M} is one of the most competitive N-body code that is publicly available for cosmologists and astrophysicists,
it has already been used  for a large number of scientific applications, and it is our hope that the current documentation will 
help the community in interpreting its outcome.
The code is publicly available on github.com under {\tt cubep3m}, and extra documentation about the structure, 
compiling and running strategy is can be found on the CITA wiki page\footnote{\tt wiki.cita.utoronto.ca/mediawiki/index.php/CubePM}.

\section*{Acknowledgements}

The development of  the {\small CUBEP$^3$M} N-body code and its utilities is continuously feeding
on discussions with users that push hard to get the science right, and we would like 
to thank them for the precious feed back they provide. 
We are also grateful Aurel Schneider for providing the halo density profiles obtained with
the AHF, and to William Watson for sharing the JUBILEE6000 power spectrum measurement. 
The CITA simulations were run on the Sunnyvale cluster at CITA.
ITI was supported by the Southeast Physics
Network (SEPNet) and the Science and Technology Facilities Council
grants ST/F002858/1 and ST/I000976/1. 
Computations for the SciNet runs were performed on the GPC, TCS and BG/Q supercomputers at the SciNet HPC Consortium. 
SciNet is funded by: the Canada Foundation for Innovation under the auspices of Compute Canada; 
the Government of Ontario; Ontario Research Fund - Research Excellence; and the University of Toronto. 
The authors acknowledge the
TeraGrid and the Texas Advanced Computing Center (TACC) at the
University of Texas at Austin (URL: http://www.tacc.utexas.edu) for
providing HPC and visualization resources that have contributed to the
research results reported within this paper. ITI also acknowledges the Partnership for Advanced
Computing in Europe (PRACE) grant 2010PA0442 which supported the code
scaling studies. ULP and JDE are supported by the NSERC of Canada, and 
VD acknowledges support by the Swiss National Science Foundation.

\appendix

\section{Example of Geometry}
\label{app:geometry}

In this appendix, we present a few of the many possible configurations for the code internal geometry, and specify the memory requirements and parameters values.
It is by no means an exhaustive list, but is meant to give a flavour of the possibilities. Most of these were already presented in the main part of this paper;
we refer to the original paper in the caption otherwise.

\begin{table*}
   \centering
     \caption{Examples of configurations and memory requirements for N-body simulations with {\small CUBEP$^3$M}, 
     extracted either from production runs or benchmark testing. 
     n$_p$ is the total number of particles per dimension, n$_{MPI}$ is the number of {\small MPI} tasks, n$_{tile}$ is the number of tile per dimension per node,
     nf$_{tile}$ is the number of fine mesh cell per tile, including the buffer, and 
     `cell/slab' is the thickness of the slab in coarse cells (used during global {\small FFTW}). 
     To estimate the memory consumption of the configuration, we assume that each node has access to 8 cores, 
     although many machines now have higher core count and/or access to hyper threading.  
     Also, we assume a standard pure N-body configuration (i.e. non-lean, no MHD), where a node can support a maximum of twice the initial particle load,
     and where particles are initially placed onto the grid with a ratio of 1 particle per 8 cells. 
     The TCS simulation suite is a series of 185 N-body independent runs that served as the {\small CLONE} catalogue for the CFHTLenS analysis 
     \citep{2012MNRAS.426.1262H, 2012MNRAS.427..146H}. They were run on the TCS super cluster at SciNet.
     The CITA864 suite was used in \citet{Vafaei10} to optimize weak lensing lensing surveys. }
      \begin{tabular}{@{} |cllcccccccccr |@{}} % Column formatting, @{} suppresses leading/trailing space
      %\toprule
      %\multicolumn{2}{c}{Item} \\
      %\cmidrule(r){1-2} % Partial rule. (r) trims the line a little bit on the right; (l) & (lr) also possible
      \hline
     n$_p$        &  n$_{MPI}$      &  n$_{tile}$              & nf$_{tile}$     & cell/slab       & mem / {\small MPI} task  (Gb) & Run Name\\
\hline
 \multirow{4}{*}{256}      &  \multirow{3}{*}{$1^3=1$}       &          2                       &   304            &          128                                   & 5.3  & \\
                                           &       &           4                       &   176           &           128                                 &       2.4            &    \\ 
                                           &         &           8                       &   112           &           128                                 &       1.9             &  SCINET256 \\ 
                                           & $2^3=8$       &          4                       &   112            &          16                                   & 0.5  & \\
\hline
 \multirow{2}{*}{432}     & \multirow{2}{*}{$3^3=27 $}   &          4                 &   192       &            16                                   &    3.2   &  CURIE432 \\
                                             &                                                    &          8                 &   120       &            16                                   &    2.6     & \\
\hline
 \multirow{2}{*}{512}     & \multirow{2}{*}{$2^3 = 8$}      &          4                      &    176            &         32                                     &       2.3    &  SCINET512\\
                                            &                                                     &          8                      &    112            &         32                                     &       1.9     &\\
\hline
 %\multirow{4}{*}{1536}     & \multirow{2}{*}{$2^3=8 $}   &4 &          240                  &            48                                   &    7.1      &\\
  %                                       &                                                    &8 &          144                        &            48                                   &    5.8   &   \\
   %                                      & \multirow{2}{*}{$4^3=64 $}   &4&          144                     &            6                                   &    1.2     & \\
   %                                      &                                                    &8 &          96                   &            6                                   &    0.9     & \\
%\hline
 \multirow{2}{*}{864}     & \multirow{2}{*}{$3^3=27 $}   &          4                 &   192       &            16                                   &    3.2   &   CITA864\\
                                            &                                                    &          8                 &   120       &            16                                   &    2.6     & \\
\hline
 \multirow{4}{*}{1024}    &      \multirow{2}{*}{$2^3=8$}&             4                &   304            &  64        & 16.1  &\\
                                          &                                                   &             8                &   176            &  64        & 13.4  & \\
                                      &    \multirow{2}{*}{$4^3 = 64$}   &               4                &   176            &      8      &     2.4  &      TCS1024 \\
                                       &                                                         &             8                &   112            &     8       & 1.8  &\\

  \hline
 \multirow{2}{*}{1536} &      \multirow{2}{*}{$4^3=64 $}   &               4          &       240                 &             12           & 7.1   &   SCINET1536\\      
                                       &                                                         &               8          &       144                 &             12            & 5.8     &      \\      

\hline
 \multirow{2}{*}{2048}   & \multirow{2}{*}{  $4^3=64 $}&                4          &      304               &            16                              &    16.1   &    \\      
                                        &                                                    &                8          &      176               &            16                              &    13.4     &  CURIE2048\\      
 %                                       &      \multirow{2}{*}{ $8^3$}     &                4          &      176               &            2                              &    2.3       & \\      
    %                                    &                                                   &                8          &      112               &            2                              &    1.9       &\\      
  \hline
 %6912                               &   $6^3=216$                            &                8         &      192                &            8                               &    18.6    &  \\      
  %\hline
 \multirow{2}{*}{4000}    &    \multirow{2}{*}{$10^3=1000$}      &                4        &      248                &            2                               &    7.9    & RANGER4000 \\      
                                            &                                                            &                8         &      148                &            2                               &    6.5    &  \\      
 \hline
 \multirow{2}{*}{5488}       &   \multirow{2}{*}{$14^3=2744$}           &                4         &      244                &             1                               &    7.6  &  RANGER5488 \\   
                                               &                                                                   &                  8      &         146             &              1                              &     6.2 &                 \\  
   \hline
 6000                                      &   $10^3=1000$           &                8         &      198                &            3                               &    21.0    & JUBILEE6000\\ 
   \hline
 6912                                      &   $12^3=1728$           &                8         &      192                &            2                               &    18.6    & BGQ6912\\ \hline
   \end{tabular}
 
   \label{tab:runs}
\end{table*}

\bibliographystyle{hapj}
\bibliography{mybib3_new}{}

\bsp

\label{lastpage}

\end{document}